\documentclass[11pt]{JHEP}

\usepackage{epsfig,bbm}
\usepackage{amssymb} 

\def\t{\theta}
\def\l{{\lambda}}

\def\d{\partial}

\def\dbz{\partial_{\bar z}}

\def\det{{\rm Det}}
\def\Tr{{\rm Tr}}
\newcommand\0{\nonumber}
\newcommand\ee{\end{eqnarray}}	 	%eqnarray
\newcommand\be{\begin{eqnarray}}
\newcommand\ba{\begin{array}}			%array
\newcommand\ea{\end{array}}

\newcommand\e{{\rm e}}

\newcommand\C{\mathbbm{C}}
\newcommand\CP{\mathbbm{CP}}
\providecommand\mathbbm{\textbf}

\preprint{SISSA/5/99/EP/FM\\\tt hep-th/9901093}

\title{Matrix String Theory and its Moduli Space}

\author{G.\ Bonelli, L.\ Bonora, F.\ Nesti, A.\ Tomasiello\\
International School for Advanced Studies (SISSA/ISAS)\\
Via Beirut 2--4, 34014 Trieste, Italy, and INFN, Sezione di Trieste\\
E-mail: \email{bonelli@sissa.it}, \email{bonora@sissa.it}, 
\email{nesti@sissa.it}, \email{tomasiel@sissa.it}}

\abstract{The correspondence between Matrix String Theory in the
strong coupling limit and IIA superstring theory can be shown by means
of the instanton solutions of the former. We construct the general
instanton solutions of Matrix String Theory which interpolate between
given initial and final string configurations. Each instanton is
characterized by a Riemann surface of genus $h$ with $n$ punctures,
which is realized as a plane curve. We study the moduli space of such
plane curves and find out that, at finite $N$, it is a
discretized version of the moduli space of Riemann surfaces: 
instead of $3h-3+n$ its complex dimensions are $2h-3+n$,
the remaining $h$ dimensions being discrete.  It turns out that as $N$
tends to infinity, these discrete dimensions become continuous, and
one recovers the full moduli space of string interaction theory.}

\begin{document}

\section{Introduction}
\label{1}

The ${\cal N}=(8,8)$ SYM on a cylindrical 2D space-time with gauge
group U(N) (hereafter referred to as Matrix String Theory (MST))
represents in the strong coupling limit a theory of type II
superstrings. This had been conjectured with various degrees of
plausibility in several papers~\cite{motl,BS,WT,DVV} (see
also~\cite{DMVV,HV,BC} and the review article~\cite{DVVr}). A step
forward for finding more compelling evidence for such a conjecture was
made in refs.~\cite{wynter,GHV,bbn1}, where it was pointed out that the
MST contains BPS instanton solutions which interpolate between
\pagebreak[3]
different initial and final string configurations via suitable
punctured Riemann surfaces. We often refer to them as {\it stringy
instantons}. In a recent paper,~\cite{bbn2}, it was shown that, in the
strong coupling limit, MST in the background of a given classical BPS
instanton solution reduces to the Green-Schwarz superstring theory
plus a decoupled Maxwell theory, and that the leading term of the
amplitude in such background is proportional to $g_s^{-\chi}$, where
$g_s$ is the string coupling constant and $\chi=2-2h-n$ is the Euler
characteristic of the Riemann surface of genus $h$ with $n$ punctures,
which characterizes the given classical solution. This is the result
one expects from perturbative string interaction theory. Needless to
say this is a strong confirmation of the abovementioned conjecture.

The results of \cite{bbn2}, although striking (and confirmed by the
present paper), were not complete. They were essentially based on a
small subset of instantons, those corresponding to the so-called
${\mathbb Z}_N$ coverings.  In this paper we intend to fill the gap by
considering any kind of coverings and constructing the corresponding
instantons. Any such instanton consists of two ingredients, a {\it
group theoretical factor} and a {\it core}. The latter corresponds to
a branched covering of the cylinder. The group theoretical factor
contains fields that satisfy WZNW-like equations with delta-function
sources.  Inside these instantons Riemann surfaces appear as branched
coverings of the base cylinder in the form of {\it plane curves},
i.e. the zero locus of polynomials of two complex variables of order
$N$ in one of them. This implies that, as we consider string
amplitudes beyond the tree level, we are bound to meet mostly singular
curves, which, in order to be classified, need to be
desingularized. One can then set out to study the moduli space of such
curves. This problem is of utmost importance, not only because one
would like to make sure that the moduli space of type IIA theory is
actually recovered within MST, but especially because one would like
to know {\it in what sense} this becomes true.

In MST there is indeed a dependence on the discrete parameter $N$
(see~\cite{suss,BFSS} for Matrix Theory): it was already noticed in
\cite{bbn2} that for finite $N$ the moduli space ensuing from MST is
only an approximation of the moduli space of type IIA superstring
theory. We are now in a position to be more precise on this issue. We
show below that, for finite $N$, the instantons of MST reproduce
exactly only the tree string amplitudes, while they cover only part of
the moduli space of higher genus Riemann surfaces with punctures.
More precisely, in a process with $n$ external string states mediated
by a Riemann surface of genus $h$, one expects $3h-3+n$ complex
moduli; at finite $N$, MST reproduces only part of these parameters,
and $h$ of them are anyhow discrete. The latter become continuous and
we recover the full moduli space of Riemann surfaces only when $N\to
\infty$.

In this paper we clarify also another issue of MST: an instanton seems
to extend at first sight over four space-time dimensions of type IIA
superstring theory. However one can show that the Riemann surfaces
corresponding to the string instantons only in particular cases are
contained in four dimensions; in general they extend over more
physical dimensions and they can actually fill up the ten dimensional
space-time of type IIA theory.

The paper is organized as follows. The \textref{2}{second section} is a rather
detailed review of previous results, as well as an overview of
problems we want to clarify in this paper. It also contains a summary
of our main results while the details of their derivation are deferred
to the following sections.  In section~\ref{3} we give the general
construction of the first ingredient of a stringy instanton, i.e. its
group theoretical factor. The last two sections are instead devoted to
a description of the second ingredient, i.e. Riemann surfaces as
branched coverings of the cylinder (section~\ref{4}) and their moduli space
(section~\ref{5}).

\section{Matrix String Theory: overview of problems and solutions}
\label{2}

\subsection{Euclidean MST and Hitchin systems}
\label{2.1}

To start with let us summarize the results of \cite{bbn1,bbn2}. MST is
a theory defined on a cylinder\footnote{In \cite{dadda,kostov} the authors
have considered related theories on the torus.} ${\cal C}$ with coordinates 
$\sigma$ and $\tau$. Its Euclidean action is
\begin{eqnarray} 
S&=&\frac{1}{\pi} \int_{\cal C} d^2w \,\Tr \Bigg(
D_w X^i D_{\bar w} X^i - \frac{1}{4g^2} F_{w\bar w}^2 -
\frac{g^2}{2}[X^i,X^j]^2 \0\\
&&~~~~~~~~~~~~~~~~~~~ +i (\theta^-_s D_{\bar w} \theta^-_s + \theta^+_c
D_w \theta^+_c) + ig \theta^T \Gamma_i [X^i,\theta] \Bigg),\label{eSYM}
\end{eqnarray}
 where we use the notation 
\begin{eqnarray}
w= \frac {1}{2} (\tau +i \sigma),\quad \bar w = \frac {1}{2} (\tau - i \sigma),
\quad\quad A_w= A_\tau-iA_\sigma ,\quad A_{\bar w}= A_\tau+iA_\sigma\,.\0
\end{eqnarray}
 
Moreover $X^i$ with $i=1,\ldots,8$ are hermitean $N\times N$ matrices
and $D_w X^i = \partial_w X^i + i[A_w, X^i]$. $F_{w\bar w}$ is the
gauge curvature.  Summation over the $i,j$ indices is understood.
$\theta$ represents 16 $N\times N$ matrices whose entries are 2D
spinors. It can be written as $\theta^T= (\theta^-_s,\theta^+_c)$,
where $\pm$ denotes the 2D chirality and $\theta^-_s,\theta^+_c$ are
spinors in the ${\bf 8_s}$ and ${\bf 8_c}$ representations of $SO(8)$,
while $^T$ represents the 2D transposition.  The matrices $\Gamma_i$
are the $16\times 16$ $SO(8)$ gamma matrices.

The action (\ref{eSYM}) has ${\cal N}=(8,8)$ supersymmetry.  In
\cite{bbn1} we singled out classical supersymmetric configurations
that preserve $(4,4)$ supersymmetry.  In this configurations the
fermions are zero, $\theta=0$, and $X^i =0$ for all $i$ except two,
for definiteness $X^i\neq 0$ for $i=1,2$. Introducing the complex
notation $X=X^1+iX^2$, $~\bar X= X^1-iX^2= X^\dagger$, the conditions
to be satisfied for such BPS configurations are
\begin{eqnarray}
&F_{w\bar w} + i g^2 [X,\bar X] =0\ \label{insteq1}\\
&D_w X=0, \qquad D_{\bar w} \bar X=0\,.\label{insteq2}
\end{eqnarray}
From a mathematical viewpoint (\ref{insteq1}), (\ref{insteq2}) can be
identified with a {\it Hitchin system}~\cite{hitchin1} on a cylinder.
Hitchin systems are defined starting from a $U(N)$ vector bundle $V$
over ${\cal C}$, associated with the fundamental representation of
$U(N)$. They consist of couples $(A,X)$ where $A$ is a gauge
connection and $X$ a section of $End V \otimes K$, where $K$ is the
canonical bundle of ${\cal C}$, which satisfy~(\ref{insteq1}) and
(\ref{insteq2}), \cite{hitchin1}. Such systems can be lifted to an
$N$-branched covering of ${\cal C}$, \cite{hitchin2, mark, WD,bochi}.  Such
lifting was the basis of all the developments~in~\cite{bbn2}.

\subsection{Construction of instanton solutions}
\label{2.2}

Each solution of~(\ref{insteq1}), (\ref{insteq2}) consists of two parts:
a branched covering of the cylinder via the relative $X$
characteristic polynomial and a group theoretical factor.  A few
explicit solutions, based on ${\mathbb Z}_N$ coverings, were presented
and discussed in \cite{bbn1,bbn2}. ${\mathbb Z}_N$ coverings are but a
very restricted set of coverings. Consequently, the
corresponding instantons are only a limited set of all possible
stringy instantons.

The aim of the present paper is to generalize the analysis of
\cite{bbn1,bbn2} by considering the most general possible coverings
and constructing the corresponding instantons. The purpose of this
section is to provide a mostly qualitative overview of the problems
involved and of the main results.

Let us start by recalling our construction of the solutions of eqs.
(\ref{insteq1}), (\ref{insteq2}). By this we mean a couple $(X, A_w)$
which are solutions of~(\ref{insteq1}), (\ref{insteq2}) and are smooth
everywhere on ${\cal C}$. We parametrize them as
\begin{eqnarray}
X = Y^{-1}M Y, \quad\quad A_w = -i Y^{-1}\d_w Y\,.\label{ansatz}
\end{eqnarray}
The group theoretical factor $Y$ takes values in the complex group
$SL(N,{\mathbb C})$ while the matrix $M$ determines the branched
covering (see below). The dependence on the Yang-Mills coupling
constant $g$ is contained in the $Y$ factor, while $M$ does not depend
on $g$. In \cite{bbn2} we have shown on several examples that $Y= Y_s
Y_d$, where $Y_d$, the dressing factor, tends to 1 in the strong
coupling limit outside the string interaction points, while $Y_s$ is a
special matrix, independent of $g$, endowed with the property that
$Y_s^{-1} M Y_s$ and $Y_s^\dagger M^\dagger (Y_s^\dagger)^{-1}$ are
simultaneously diagonalizable. The construction of $Y_s$ and $Y_d$ in
the general case is rather subtle. One first diagonalizes $M$ by means
of a matrix $S$ of $SL(N,{\mathbb C})$. Then one introduces a matrix
$K$ such that $KS=U$ is unitary. As it turns out, $K$ may have
singularities at the points of ${\cal C}$ where any two eigenvalues of
$M$ coincide: these correspond to the branch points of the spectral
covering ($K$ is also allowed to diverge in a prescribed way for
$w=\pm\infty$, but we disregard this issue for the time
being). Therefore $KMK^{-1}$ is in general singular at these
points. We therefore introduce into the game a new matrix $L$, with
the purpose of canceling the singularities of $KMK^{-1}$ in such a way
that $LKMK^{-1}L^{-1}$ is smooth and satisfies
(\ref{insteq1}),~(\ref{insteq2}). In order for this to be true the
entries of $L$ must satisfy equations of the WZNW type with
delta-function-type sources at the branch points. By construction $K$
is independent of $g$ while $L$ does depend on $g$. We will show that
in fact $L\to 1$ as $g\to \infty$. We therefore see that $K^{-1}$
plays the role of $Y_s$ and $L^{-1}$ is to be identified with the
dressing factor $Y_d$, so that $LK=Y^{-1}$.

We will deal with the general construction of $Y_s$ and $Y_d$ in
detail in section~\ref{3}. Here we would like to stress the double `miracle'
of the above construction: we construct an everywhere smooth solution
by means of two non smooth matrices $K$ and $L$, which are such that
on the one hand $L\to 1$ as $g\to \infty$ and on the other hand $K$
form with $S$ a unitary matrix $U=KS$. This can be adequately
appreciated in the light of ref.~\cite{bbn2}. It was shown there that,
thanks to these properties, in the strong coupling limit, we can get
entirely rid of the non-diagonal background part. In order to deal
with the singularities that are exposed as the dressing factor tends
to 1, one cuts out a small disc around any string interaction point
(i.e. one introduces a regulator), defines the theory on the cylinder
minus such discs (where, in the strong coupling limit, $Y_d=1$) and
get rid of the $U$ factor by means of a gauge transformation.  At this
point the analysis of \cite{bbn2} can be conveniently carried out and,
eventually, the regulator removed.

\subsection{Riemann surfaces as branched coverings of the cylinder} 
\label{2.3}

The second ingredient of an instanton solution is a branched covering
of the cylinder.  In dealing with branched coverings it is however
more convenient to map, by the standard mapping $z = e^{\bar
w}$, the infinite cylinder ${\cal C}$ to the punctured complex
$z$-plane ${\mathbb C}\setminus\{0\}$, i.e.  ${\mathbb C}^*$. Very
often we will refer to it as the Riemann sphere $\CP^1$ (with two
punctures).  

Let us consider the polynomial
\vspace*{-.5em}
$$
P_X(y)=\det (y - X)=y^N+\sum_{i=0}^{N-1}y^ia_i\,,
$$
where $y$ is a complex indeterminate.  Due to~(\ref{insteq2}), we have
$\dbz a_i=0$ which means that the set of functions $\{a_i\}$ are
analytic on the complex plane,\footnote{Notice the change of convention
with respect to \cite{bbn1,bbn2}, which amounts to the exchange $z
\leftrightarrow \bar z$. This is in order to refer to the $a_i$'s as
analytic functions, rather than antianalytic.} although they are
allowed to have poles at $z=0$ and $z=\infty$.  Therefore the equation
\be 
P_X(y)=0\label{spec} 
\ee 
identifies in the $(z,y)$ space (i.e. ${\mathbb C}^*\times {\mathbb
C}$, but often in the following for simplicity we replace it 
with the affine space ${\mathbb
C}^2$ ) a Riemann surface $\Sigma$, which is an N-sheeted branched
covering of the cylinder. For later use we recall that eq. $P_X(y)=0$
is tantamount to considering the matrix equation
\vspace*{-.5em}
\be 
X^N+a_{N-1}X^{N-1}+\cdots +a_0=0\,.\label{spectr} 
\ee 
The explicit form of the covering map is given by the set
$\{x^{(1)}(z),\dots,x^{(N)}(z)\}$ of eigenvalues of $X$. Each
eigenvalue spans a sheet. The projection map to the base cylinder
${\cal C}$ will be denoted $\pi:\Sigma \to {\cal C}$. The points where
two or more eigenvalues coincide are called branch points. The
identification cuts in the sheets start or end at these points. We
would like to warn the reader that with the term `branch point' we
denote, in general, both the points on the covering and their image in
${\cal C}$ under the projection $\pi$. The context should make clear
which exactly we refer to.
 
A diagonalizable matrix $M$, solution of eq.~(\ref{spectr}),
can always be cast in canonical form
\be
M=\left(\matrix{-a_{N-1}& -a_{N-2}& \ldots & \ldots  & -a_0\cr
		  1      & 0       & \ldots & \ldots  & 0  \cr
		  0       & 1       &  0  & \ldots  & 0 \cr
		  \ldots	   & \ldots     &  \ldots& \ldots  & 
		  \ldots \cr		
                  0	    & 0       & \ldots & 1    & 0 \cr
}\right).\label{M}
\ee
The branched covering structure is completely encoded in the $\{a_i\}$
analytic functions and we have already stressed that it is independent
of the coupling $g$.

\pagebreak[3]

\subsection{String interpretation}
\label{2.4}

This is a good point for reviewing the string theory interpretation of
solutions $(X,A_w)$ of~(\ref{insteq1}),~(\ref{insteq2}) that we
presented in \cite{bbn1,bbn2}, while enriching it with new remarks
pertinent to the present paper. We recall that we interpret the
Riemann surface defined by the relevant branched covering of the
cylinder as the classical carrier of a string process. The branch
points at $z\neq 0,\infty$ represent joining and splitting processes
of the string. Generically, when the branch point is simple, we have
the joining of two strings to form a unique string or the splitting of
one string into two. We may also have multiple branch
points\footnote{The multiplicity or ramification index of a branch
point is defined as number of sheets which come together at that
point, minus one; therefore branch points that involve only two sheets
are called simple.}  in which more then two incoming or outgoing
strings are involved. However in this paper, contrary to
\cite{bbn1,bbn2}, the emphasis will be on simple branch
points.\footnote{This is to simplify the exposition. Multiple branch
points at $z\neq 0,\infty$ can also be analyzed.  They define anyhow
lower dimensional subspaces of the moduli space and can be obtained as 
limiting cases of simple branch points.}
 
The inverse images under $\pi$ of $z=0$, $z=\infty$ are punctures in
$\Sigma$ with a definite string interpretation: they represent
incoming and outgoing strings, respectively. More precisely they
represent the points where incoming strings enter (outgoing strings
leave) the process represented by the Riemann surface $\Sigma$.
 
It has to be kept in mind that, in MST, the counterimages of $z=0$ and
$z=\infty$ are distinguished points with an associated physical
meaning. This is to be contrasted with the usual mathematical
treatment of branched coverings of $\CP^1$, where these points do
not play any particular role.  This remark will become extremely
important below, in connection with the discussion about moduli space.

Let us discuss further properties of the punctures corresponding to
$z=0$ (an analogous discussion holds for $z=\infty$), see for example
\cite{bbn1}.  The counterimages of $z=0$ by $\pi$ may be $N$ distinct
points, i.e. the solutions of the algebraic equation~(\ref{spec}) at
$z=0$ may be all distinct. In such a case we say we have $N$ small
incoming strings (of length 1 each). However, in general, the inverse
image of $z=0$ may contain several branch points
$P_1^{in},\ldots,P_s^{in}$, with multiplicity $l_1-1,\ldots,l_s-1$,
respectively (if $z=0$ is a singular point of the eq.~(\ref{spec}) it
has to be desingularized first, see below). In this case the process
represented by $\Sigma$ involve $s$ incoming strings of length $l_1,
\ldots, l_s$, respectively. The physical interpretation of the string
length has been given in \cite{DVV}. In the framework of the
light-cone quantization of type IIA superstring, the string length is
identified with the momentum component $p^+= p^9+p^0$ of the string in
suitably normalized units. Here $0,9$ are of course the time and
longitudinal direction of the ambient space, %which do 
not explicitly
appearing  in~(\ref{eSYM}).
 
Let us summarize the string interpretation of the solutions of
(\ref{insteq1}),~(\ref{insteq2}). Each such solution is characterized by
a punctured Riemann surface realized as a branched covering of the
cylinder. The punctures represent the sites where strings enter or
leave the interaction process. The length of each string is associated
to its $p^+$ momentum component.  This is the picture of MST at strong
coupling. At finite coupling $g$ the string interpretation of the
instantons persists, but the dressing factor $Y_d$ has the effect of
blurring it by smearing the string interactions and so on.

Now, the question arises of how to transform this scenario into an
effective calculational tool. One definite suggestion has been made in
\cite{bbn2}.  One considers~(\ref{eSYM}) and expands it about a
classical solution which spans a given string process (Riemann surface
with punctures), then takes the $g\to \infty$ limit. It turns out that
the dressing factor of the instanton disappears and the unitary factor
can be gauged away. The strong coupling limit turns out to be well
defined and, practically, to the lowest order in $1/g$, the background
fields disappear, except for the notion of the covering surface
$\Sigma$. This can be seen by splitting all the matrix fields into
Cartan (diagonal) and non-Cartan modes. A suitable gauge fixing makes
it clear that the non-Cartan part can be integrated out, while the
Cartan modes can be interpreted as fields on $\Sigma$.  The final
upshot is that MST at strong coupling reduces to the Green-Schwarz
superstring theory plus a decoupled Maxwell theory, and that the
leading term of the amplitude in such background is proportional to
$g_s^{-\chi}$, where $g_s=g^{-1}$ is the string coupling constant and
$\chi=2-2h-n$ is the Euler characteristic of $\Sigma$ if its genus is
$h$ and its punctures are $n$.

Let us denote, as in \cite{bbn2}, the fields in $\Sigma$ with a tilde:
$\tilde x, \tilde a, \tilde \theta, \tilde c$ are the small
oscillations of $X, A_w, \theta$ and the ghost $c$, respectively,
which survive in the strong coupling limit. Then, in order to describe
a complete string process, we can introduce the vertex operators
$V_1,\ldots , V_n$ corresponding to $n$ incoming and outgoing strings,
expressed in terms of $\tilde x, \tilde \theta$, possibly the
superstring reparametrization ghosts, and of the string transverse
momenta, and insert them into the path integral. The amplitude (in the
strong coupling limit) will schematically be:
\be
\langle V_1,\ldots,V_n\rangle_h= g_s^{-\chi}\int_{{\cal M}^{(h,n)}_{N}}dm~ 
\int {\cal D}[\tilde{x},\tilde{\theta},\tilde{a},\tilde{c}] J_{b/f}J_{C/nC} 
V_1\ldots V_n\e^{-S_{GS}-S_{Maxwell}}\,,
\ee
where most of the symbols are the same as in \cite{bbn2}. Here we have
singled out the integration over ${{\cal M}^{(h,n)}_{N}}$, namely over
all distinct instantons which underlie the given string process for
fixed $N$, that is to say with assigned incoming and outgoing strings
and string interactions. In ordinary string interaction theory ${{\cal
M}^{(h,n)}}$ is nothing but the moduli space of Riemann surfaces of
genus $h$ with $n$ punctures, a complex space of dimension
$3h-3+n$. What actually ${{\cal M}^{(h,n)}_{N}}$ is in MST is the main
subject of the present paper.

\subsection{Plane curves and moduli space}
\label{2.5}

So far we have tacitly given for granted that any Riemann surface with
$h$ handles and $n$ punctures can be represented as a branched
covering of ${\CP^1}$. This point has to be carefully handled.

Let us start from eq.~(\ref{spec}). This is the equation of a curve in
the affine space ${\mathbb C}^2$ spanned by the two complex
coordinates $y$ and $z$. A curve embedded in a two-dimensional affine
space is called a {\it plane curve}. Therefore MST dynamically
engenders Riemann surfaces in the form of plane curves.
\pagebreak[3]

An important role in the following is played by singular plane curves.
If $P(y,z)=0$ is the polynomial equation of the curve, a singular
point is a point where $P(y,z)= \d_y P(y,z) = \d_z P(y,z)=0$.  When no
such points are present the curve is smooth. However this can happen
only when its genus is ${1\over 2} (d-1)(d-2)$, where $d$ is the
degree of the curves, i.e. the degree of the polynomial $P(y,z)$: we
will see later on that in our case the degree is $N$, see
section~\ref{4}. We can lower the genus of the plane curve, while
keeping the degree constant, by allowing for singularities.  This
means two important things: first, for finite $N$ there exists an
upper bound (${1\over 2}(N-1)(N-2)$) on the genus of the Riemann
surfaces which define the {\it core} of the stringy instantons;
second, far from discarding singular curves, as one would be tempted
to do as a first approach, we have to take them into account, they are
bound to fill up most of the moduli space of plane curves. As we will
see, singular curves are a happy occurrence, not a nuisance.

The formulation of plane curves in terms of the variables $y$ and $z$
is at times ambiguous, especially in connection with singularities. A
possible way to resolve such ambiguities is to embed them in the
projective space ${\CP^2}$ by introducing three homogeneous
coordinates $x_0,x_1,x_2$. Simply set $z=x_1/x_0, y= x_2/x_0$ in
eq.~(\ref{spec}), and multiply by a suitable power of $x_0$. The new
equation refers to the same curve embedded in ${\CP^2}$. This is a
`cleaner' representation and the one we use in the following.
Singular curves can be desingularized, for example by (repeatedly)
blowing up the singular points until we reach smooth
configurations. However the smooth curves one obtains this way are not
plane curves. In order to be represented as algebraic curves they need
to be embedded in a larger affine or projective space. It is
well-known that any compact Riemann surface can be represented as a
smooth algebraic curve embedded in ${\CP^3}$. However, if we try to
project it to ${\CP^2}$, i.e. to represent it as a plane curve, we
are bound to produce singularities.  The lesson we learn is that we
have to count mostly on singular plane curves in order to reproduce
the moduli space of Riemann surfaces which is needed in string
interaction theory. There is nothing arbitrary about this: all the
information we need to reproduce string theory (topology and moduli)
is contained in the singular curve: if we know the singular curve we
can reproduce a smooth counterpart with a standard algorithm.  Once
this point about plane curves is clarified, we will not talk about
singular or regular plane curves but simply about plane curves.  Plane
curves will be discussed in detail in section~\ref{3}.

At this point it would seem that we are done: a theorem by Clebsh,
\cite{brie}, guarantees that any (compact) algebraic curve is
birationally equivalent to a plane algebraic curve which has at most
ordinary double points as singularities. However this is too
simplistic.  Beside the upper bound for the genus we have mentioned
above, we should remember that in our case we do not have to do with
compact Riemann surfaces, but with Riemann surfaces with a certain
number of punctures. Therefore the above theorem is not
conclusive. Actually we will find out in section~\ref{4} that the
presence of punctures on the Riemann surface entails the consequence
that the {\it moduli space of plane curves of genus $h$ with $n$
punctures is a discretized version of the the moduli space of genus
$h$ Riemann surfaces with $n$ punctures}, whose complex dimension is
$3g-3 +n$. A good parametrization of the moduli space, fit for string
interaction theory, is provided by Mandelstam's variables
\cite{mand,gidwol,dhofo,GSW}.  By making a comparison with
this parametrization, we will find out that {\it $h$ of the
Mandelstam complex parameters are actually discrete for the plane
curves that appear in MST}.

This point is rather technical and we only have a technical
explanation for it (section~\ref{4}). Its origin can be briefly described as
follows. The coordinate $z$ we have introduced above, can be naturally
regarded as a meromorphic function on a given plane curve (it is a
realization of the projection $\pi: \Sigma \to {\cal C}$). The
counterimages of $z=0$ and $z=\infty$ form a principal divisor in
$\Sigma$. This entails, by Abel's theorem, $h$ discretizing conditions
on the parameters describing the plane curve. A detailed analysis
shows that this imposes $h$ of the Mandelstam parameters to be
discrete.  A confirmation of this result comes from an estimate of the
moduli space of stringy instantons. Since in the $Y$ factor there no free
parameters, the moduli space of stringy instantons must coincide with the
free parameters contained in $M$, i.e. with the moduli space of plane
curves. The estimate carried out in section~\ref{5} confirms the above
evaluation of the continuous dimension of the moduli space of the
latter.

{\it However when $N\to \infty$ these discrete parameters become
continuous and, in addition, the upper limit on the genus we mentioned
above becomes ineffective}. Therefore for large $N$ MST recovers the
full moduli space of string theory. We recall that, for finite $N$,
also the $p^+$ components of the momenta of the incoming and outgoing
strings are discrete and continuity is recovered only for $N \to
\infty$. Therefore, {\it a complete description of string interaction
theory is truly achieved by MST only in the large $N$ limit.}  It is
nevertheless remarkable that genus 0 processes (with discrete $p^+$
components of the external momenta) are exactly described by MST also
for finite (but large enough) $N$.
 
\subsection{Singularities and space-time dimensions}
\label{2.6}

Singularities of plane curves provide a striking suggestion of how to
resolve an unsatisfactory aspect of the correspondence between MST and
string theory. In~(\ref{eSYM}) we see eight ambient space dimensions
and two world-sheet dimensions of the cylinder. We should now
remember that the correspondence MST ---string theory is established
in the light-cone gauge. Therefore the two world-sheet dimensions
are nothing but representatives of the time and longitudinal
dimensions, denoted 0 and 9, which bring the total of physical
dimensions to ten. Now, stringy instantons seem to span four out of
these ten dimensions. In other words it would seem that MST at strong
coupling can only describe four-dimensional string processes.
However this is strictly true only for processes which are mediated by
smooth plane curves. As we have pointed out, singular curves become
smooth if we enlarge the space where they are embedded. For example,
curves in ${\CP^2}$ with nodes only, can be smoothed out by
embedding them in ${\CP^3}$, that is by adding two dimensions, and
so on. We interpret this by saying that the corresponding string
process extend over six (instead of four) dimensions. It is not
difficult to envisage processes that extend over ten dimensions. One
can phrase it as follows: all these high dimensional processes are
squeezed to four dimensions in order to fit into the instantons of the
2d field theory~(\ref{eSYM}), and this projection gives rise to
singularities.

\section{The unitary and dressing factors}
\label{3}

As explained in the previous section, the MST instantons consist of
two pieces: a group theory factor $Y$ and a branched covering (plane
curve) parametrized by the matrix $M$. In section~\ref{4} and~\ref{5}
we will discuss the latter. Let us now concentrate on the former. In
the derivation of the results of ref.~\cite{bbn2} it is of paramount
importance that $Y$ splits according to $Y= Y_s Y_d$, where $Y_d\to 1$
as $g\to \infty$ outside the string interaction points (branch points
at $z\neq 0,\infty$), and $Y_s$ is a matrix independent of $g$ such
that $Y_s^{-1} M Y_s$ and $Y_s^\dagger M^\dagger (Y_s^\dagger)^{-1}$
commute.

In this section we prove that the above structure of the $Y$ factor
holds for a general covering. In other words, we start from a general
matrix $M$~(\ref{M}), construct the corresponding $Y$ factor and show
that it satisfies the requirements we have just mentioned.

\subsection{The unitary factor}
\label{3.1}

The factor $Y_s$ can be constructed as follows. It is well-known,
\cite{wynter}, that the matrix $M$ can be diagonalized
\begin{eqnarray}
M = S D S^{-1}, \quad\quad D = {\rm Diag}(\lambda_1,\ldots, \lambda_N)
\end{eqnarray}
by means of the following matrix $S\in SL(N,{\mathbb C})$:
\be
S= \Delta^{-{1\over N}}\left(
	\matrix{\l_1^{N-1}&\l_2^{N-1}& \ldots & \ldots &\l_N^{N-1}\cr
		\l_1^{N-2}&\l_2^{N-2}& \ldots & \ldots  &\l_N^{N-2}\cr
		\ldots	& \ldots  & \ldots & \ldots  &  \ldots \cr
                1	& 1	  & \ldots & \ldots  & 1 \cr}\right),\label{S}
\ee
where 
\be
\Delta = \prod_{1\leq i<j\leq N}(\l_i - \l_j)\,.\label{Delta}
\ee
We now introduce a matrix $K$ such that $U=KS$ is unitary. If such a
matrix exists than $ KMK^{-1}= UDU^{-1}$ and $(K^\dagger)^{-1}
M^\dagger K^\dagger= U D^\dagger U^{-1}$ do commute. Then $K$ can play
the role of $Y_s^{-1}$.  We refer to $U=KS$ as the {\it unitary
factor} in the construction of the background solution $X$.

One such $K$ can be constructed with the Gram-Schmidt procedure. 
The result is the following upper triangular matrix belonging to 
$SL(N,{\mathbb C})$:
\be
K=\left(\matrix{k_{11}&k_{12} & \ldots & \ldots  & k_{1N}\cr
		   0     & k_{22}       & \ldots & \ldots  &k_{2N}  \cr
		  \ldots	   & \ldots     &  \ldots& \ldots  & 
		  \ldots \cr		
                  0	    & 0       & \ldots & \ldots    & k_{NN} \cr
}\right),\label{K}
\ee
where
\be
k_{pp} &=& \sqrt{\frac {Q^{(N-p)}}{Q^{(N-p+1)}}} |\Delta|^{1\over N},\quad\quad
\quad\quad 1\leq p\leq N\0\\
k_{pq} &=& \frac{Q^{(N-p+1)}_{N-q,N-p}} {\sqrt{Q^{(N-p)}Q^{(N-p+1)}}}
|\Delta|^{1\over N}, \quad\quad 1\leq p<q\leq N-1\,.\label{kpq}
\ee
The symbols in the previous formulae have the following meaning.
For $2\leq k\leq N$ we set
\be
Q^{(k)} &=& \sum_{1\leq j_1<j_2<\ldots<j_k\leq N} \left(\prod_{1\leq l\leq k-1}
|\l_{j_l}- \l_{j_{l+1}}|^2|\l_{j_l}- \l_{j_{l+2}}|^2\ldots
|\l_{j_l}- \l_{j_k}|^2\right),\0\\[1.5ex]
Q^{(k)}_{n,m}&=& (-1)^{n+m}\!\!\!\!\sum_{1\leq j_1<j_2<\ldots<j_{k-1}\leq N}
\!\!\!\!S_{k-1-n}(\l_{j_1}, \l_{j_2},\ldots, \l_{j_{k-1}})
S_{k-1-m}(\bar\l_{j_1}, \bar\l_{j_2},\ldots,\bar \l_{j_{k-1}})\times\0\\[-.2ex]
&&
\times\!\left(\delta_{k,2}+\!\!\!\!\prod_{1\leq l\leq k-2}\!\!\!\!
|\l_{j_l}- \l_{j_{l+1}}|^2 |\l_{j_l}- \l_{j_{l+2}}|^2 \!\ldots
|\l_{j_l}- \l_{j_{k-1}}|^2\right), \quad 0\leq n\leq m\leq N-1\0
\ee
where $S_p(x_1,\ldots,x_s)$ denotes the elementary symmetric polynomial
of order $p$ in $x_1,\ldots,x_s$ ($s\geq p$):
\be
S_p(x_1,\ldots,x_s)= \sum_{1\leq j_1<j_2<\ldots<j_p\leq s}x_{j_1}\ldots
x_{j_p},\quad\quad S_0 =1\,.\label{Sp}
\ee
Moreover, by definiton, $Q^{(0)}=1, Q^{(1)}= N$.
Notice that
\be
Q^{(N)} = |\Delta|^2,\qquad k_{11}k_{22}\ldots k_{NN}=1\,. \0
\ee
The $Q^{(k)}_{n,m}$ are homogeneous polynomials of order $\frac{k(k-1)}{2}-n$
in the variables $\l_i$ and of order $\frac{k(k-1)}{2}-m$ in the complex
conjugates $\bar \l_i$. $Q^{(k)}$ are homogeneous polynomials of order
$\frac{k(k-1)}{2}$ in both $\l_i$ and $\bar \l_i$. 

For example, in the case $N=3$, the matrix $K$ is given by
\be
K=\left(\matrix{\frac{\sqrt {\sum_{i<j}|\l_i-\l_j|^2}}{|\Delta|^{2\over 3}}&
- \frac{\sum_{i<j}(\l_i-\l_j)|\l_i-\l_j|^2}{{\sqrt {\sum_{i<j}|\l_i-\l_j|^2}}
|\Delta|^{2\over 3}}&\frac{\sum_{i<j}\l_i\l_j|\l_i-\l_j|^2}
{{\sqrt {\sum_{i<j}|\l_i-\l_j|^2}}|\Delta|^{2\over 3}} \cr
		  0&\sqrt{\frac{3}{\sum_{i<j}|\l_i-\l_j|^2}}|\Delta|^{1\over 3}&
 - \frac{\sum_i\l_i}{\sqrt{3\sum_{i<j}|\l_i-\l_j|^2}}|\Delta|^{1\over 3}\cr
0   & 0     & \frac{|\Delta|^{1\over 3}}{\sqrt 3} \cr		
}\right), \quad 1\leq i,j\leq 3\,.\0
\ee

This completes the construction of $K$. We remark that generally the
entries of $U=KS$ contain as a factor some fractional power of
$|\Delta|$. Therefore they may vanish or diverge with some fractional
power whenever two of the eigenvalues of $M$ coincide. This
corresponds to a simple branch point in the spectral covering, as we
have seen above.\footnote{The entries of $K$ contains other factors,
beside $\Delta$, that may vanish when more than two eigenvalues
coincide. This corresponds to multiple branch points, which we
disregard in this paper.} Outside these points the unitary factor $U$
is smooth. It is therefore justified to get rid of it by a gauge
transformation, as we have done in \cite{bbn2}.

\subsection{The dressing factor}
\label{3.2}

Let us come now to the $Y_d$ factor. We have just noted that the
entries of $K$ are generically singular whenever two eigenvalues of
$M$ coincide, that is at the site of a branch point of the
covering. Then $KMK^{-1}$ shares the same singularities and it is not
a satisfactory ansatz for our solution $X$ of
(\ref{insteq1}),~(\ref{insteq2}), which we want to be everywhere
smooth (except perhaps at $w=\pm \infty$). To this end we introduce a
new matrix $L$ with the requirement that $LKMK^{-1}L^{-1}$ is smooth
and is the desired solution $X$ of
(\ref{insteq1})~,(\ref{insteq2}). While $K$ is independent of $g$, $L$
will depend on $g$. $L^{-1}$ is our candidate for the dressing factor
$Y_d$.

Since $L$ has to smooth out the singularities of $K$, it is enough to take 
for it an upper triangular matrix belonging to $SL(N,{\mathbb C})$. A possible 
parametrization for $L$ is the following
\be 
L =\left(\matrix{e^{u_1}&e^{u_1}\psi_{12} &e^{u_1}\psi_{13} &\ldots 
&e^{u_1} \psi_{1N}\cr
		   0     & e^{u_2-u_1} &e^{u_2-u_1}\psi_{23} & \ldots 
&e^{u_2-u_1}\psi_{2N}  \cr
		  \ldots	   & \ldots     &  \ldots& \ldots  & 
		  \ldots \cr		
                  0	    & 0       & \ldots &e^{u_{N-1}-u_{N-2}}    &
e^{u_{N-1}-u_{N-2}}     \psi_{N-1N} \cr
0	    & 0       & \ldots & 0&e^{-u_{N-1}} \cr
}\right).\label{L}
\ee
The fields $u$ and $\psi$ have to satisfy certain differential equations 
in order to comply with the requirements. We just plug the ansatz for $X$ (and
the connection) into~(\ref{insteq1}) and work out the relevant equations. 
We will not write them down here. A few examples were given in
\cite{bbn1,bbn2}. They are all equations of the WZNW type
and can be cast in the general form
\be
\d_w\d_{\bar w}\phi + ... \sim \d_w\d_{\bar w} \ln |\Delta|= {\pi}
\frac{\d \Delta}{\d w}\frac{\d \bar \Delta}{\d{\bar w}}\delta({\Delta})\,,
\label{dressing}
\ee
where $\phi$ denotes any field $u$ or $\psi$, while dots represent all the
other terms, which are irrelevant in the cancellation of
singularities. Let us refer to these equations as the `dressing
equations'.  On the right-hand side we see the typical
delta-function type source which characterizes them. The sources are
point-like and located at the zeroes of $\Delta$,
that is at the branch points of the covering.

Naturally the solution $X$ exists with the required properties only if
the `dressing equations' admit solutions that vanish at $w=\pm
\infty$. To our best knowledge, not much is known in the literature
concerning the existence of such solutions. Some simple cases were
discussed in \cite{bbn2}, where we also presented a few numerical
solutions. We deem these sufficient for us to assume that the
`dressing equations' do admit solutions that vanish at $w=\pm
\infty$. Once one assumes this, it is rather easy to argue, on a
completely general ground, that in the strong coupling limit, $g \to
\infty$, such solutions vanish outside the zeroes of the discriminant.

The argument goes as follows. Consider a candidate solution of
(\ref{insteq1}) in which $u=\psi=0$ outside the zeroes of the
discriminant. Then, there, $L=1$, and $X= KM K^{-1}$. As we have
already noticed in section~\ref{2}, in such a situation $[X, \bar X]=0$,
since both $X$ and $\bar X$ are simultaneously diagonalized by the
matrix $U=KS$. This is most welcome since, if $[X, \bar X]$ were not
to vanish (outside the zeroes of the discriminant), it would be
impossible to have any finite solution of~(\ref{insteq1}). Next, we
have to show that also $F_{w\bar w}$ vanishes in the same region when
$L=1$. In fact when $L=1$, 
\be
A_w = - i K\d_w K^{-1}=-i (KSS^{-1})\d_w (SS^{-1}K^{-1})= -i U (\d_w +
\tilde A_w)U^{-1}\,,\0
\ee
where $ \tilde A_w = S^{-1}\d_w S$. But $\d_w S\equiv 0$ due to
holomorphicity of the eigenvalues of $M$. In conclusion
(\ref{insteq1}) is identically satisfied by the ansatz $L=1$ outside
the zeroes of the discriminant. Since the solutions are uniquely
determined by their boundary conditions, we can conclude that, as
$g\to \infty$, the only solution of the dressing equations outside the
zeroes of the discriminant, is the identically vanishing solution. We
infer from this argument that the solutions of the dressing equations
for large $g$ are concentrated around the branch points and become
more and more spiky as $g$ grows larger and larger. Therefore the
matrix $L$ is just the dressing factor $Y_d$.

This conclusion, as pointed out in section~\ref{2}, justifies our
approach in \cite{bbn2}. We warn the reader, however, that the above
argument is not entirely satisfactory, for it assumes that, as $g\to
\infty$, the only possibility is $[X, \bar X]=0$ and $F_{w\bar
w}=0$. One can envisage other types of solutions. For instance, let
$A_w$ and $X$ be any $g$-dependent solution considered so far. Then
call $A_w^{(1)}, X^{(1)}$ their value at $g=1$. The latter satisfy
(\ref{insteq1}),~(\ref{insteq2}) for $g=1$. Now set $X_g=
X^{(1)}/g$. Then $ A_w^{(1)}, X_g$ satisfy the same equation, but, as
$g\to \infty$, $X_g\to 0$ while $F_{w \bar w}^{(1)}\neq 0$. Therefore
the eigenvalues of $X_g$ for $g\to \infty$ do not describe any 
asymptotic string configuration.  

We call {\it stringy instantons} those solutions of the Hitchin equations
(\ref{insteq1}),~(\ref{insteq2}) for which at $g=\infty$ 
we have both $[X, \bar X]=0$ and $F_{w\bar w}=0$. In this paper we consider 
only this kind of solutions.

\section{Branched coverings of Riemann surfaces and plane curves}
\label{4} 

This section is devoted to the detailed analysis of the stringy
instanton {\it core}, i.e. of branched coverings of the Riemann sphere.
The latter appear in MST as solutions of affine equations 
\be
P(y,z)\equiv \sum_{p,q} a_{p,q} y^qz^p=0, \quad\quad 
(y,z) \in {\mathbb C}^2\,,\label{pol}
\ee
where $P$ is a polynomial of degree $N$.  Actually this does not
follow immediately from what we said in section~\ref{2.2}. From
eq.~(\ref{spec}) it follows that $P_X$ has degree $N$ in $y$, but
the $a_i(z)$'s could be any analytic functions on the punctured
Riemann sphere. This means that they could be expressed by means of
Laurent series in $z$. However in order to preserve the string
interpretation we will limit ourselves to $a_i(z)$'s which are Laurent
polynomials and, for simplicity, in this paper we will explicitly
consider only $a_i(z)$'s which are polynomials in $z$ in such a way
that $P(y,z)$ has overall degree $N$.

The locus in ${\mathbb C}^2$ of the solutions $(y,z)$ of~(\ref{pol})
is a plane curve.  The independent non-vanishing coefficients
$a_{p,q}$ can be varied without changing, in general, the topological
type $(h,n)$ of the curve. They are the {\it moduli} of the plane
curve. Counting them is an exercise we have to do in order to see
whether the moduli space of MST coincides with the moduli space of IIA
superstring theory.

The literature on plane curves, and, more generally, on algebraic
curves is vast (see for instance \cite{brie,GH,kirwan}), and we will be using many well-known
results.  However one should bear in mind a peculiarity of our problem
which is not usually considered in the textbooks on the subject. This
is the question of punctures, which has already been introduced in
section~\ref{2.4}. Let us discuss it now in more detail.

\subsection{Punctures on plane curves} 
\label{4.1}

Punctures are the sites on the embedded Riemann surface, that is on
the corresponding plane curve, where the incoming strings enter and
the outgoing strings exit.\footnote{In all the figures below we show
the incoming and outgoing strings not as punctures, but as macroscopic
strings in order to stress their different lengths.} They are the
counterimages by $\pi$ of $z=0$ and $z=\infty$, respectively.  If any
such point on the plane curve is a branch point of multiplicity $l-1$,
then the corresponding incoming or outgoing string has length $l$.
Incidentally, since eventually we want to take the large $N$ limit, we
are especially interested in the case when $l$ is comparable with $N$.
In the ordinary treatment of compact Riemann surfaces, if these points
are regular, they are in no way special and must be considered on the
same ground as all the other regular points (this can be seen for
example by using projective coordinates). In our approach, this is not
allowed.  As we have already pointed out in section~\ref{2} the length
of an incoming or outgoing string is interpreted as the $+$ component
of the momentum in the light-cone framework. Therefore, non only the
locations of the branch points in the inverse image of $z=0,\infty$,
but also their multiplicities have a precise physical meaning.  Two
processes that differ by these multiplicities must be kept distinct,
even if, say, the topological type is the same.

Let us see an example. Suppose $y=y_1$ is a branch point of
multiplicity $l-1$ in the counterimage of $z=0$. This means that we
have $l$ roots of~(\ref{spec}). For example, $y^{(i)}\sim y_1+\eta^i
z^{1/l}$, $i=0,\ldots,l-1$ and $\eta= exp(2\pi i/l)$. In other words $l$ sheets
of the covering join along a cut starting at $y_1$.  The counterimage
of a circle around $z=0$ in the $z$-plane contains a curve around
$y=y_1$ on the covering that closes after crossing the cut $l$ times,
i.e. we have an incoming string of length $l$.  Therefore an easy rule
to compute the length of an asymptotic string at a branch point in the
inverse image of $z=0$ is to count the number of sheets that meet
there. Alternatively such length can be seen as the period of the
differential $d\ln z$ around the point $y=y_1$ of the covering. In
fact $y-y_1$ is a good coordinate near $y_1$ and $d\ln z = k~ d\ln
(y-y_1)$. The same conclusion can be drawn if the roots are like
$y^{(i)}\sim y_1+\eta^i z^{j/l}$, where $j$ and $l$ are relatively
prime integers.  A similar discussion can be carried out for the
counterimages of $z=\infty$ as well.

At this point it is convenient to mention the concept of {\it tameness}. 
Tameness was introduced by C.T.Simpson \cite{simpson,biswas}. It
means the following.  Eq.~(\ref{spec}) is an algebraic equation of
order $N$ in $y$; if one solves for $y$ one gets $N$ (possibly
coincident) roots as functions of $z$, whose behaviour near $z=0$ and
$z=\infty$ will be dominated by some (in general, fractional) power of
$z$. We say that the curve (more properly, the corresponding Hitchin
system) is tame if these roots have at most a simple pole at
$z=0,\infty$. Tameness guarantees the existence of a well-behaved
bundle metric in the $V$ bundle mentioned in section~\ref{3.1}.

If tameness is necessary from a mathematical point of view, there does
not seem to be any physical motivation for it. Let us remark that a
process is tame if the roots at the punctures behave as $y\sim {\rm
const}+ \zeta^{j/l}$, where $\zeta$ is the appropriate local
coordinate at $z=0,\infty$ and $|j/l|\leq 1$.  Now suppose that one of
the roots of~(\ref{spec}) is not branched at $y_1$ which lies in the
counterimage of $z=0$, and behaves like $y_1+z^2$. This process is not
tame but it represents an allowed string configuration. Anyhow, in the light
of the large $N$ limit, the problem of untameness becomes somewhat irrelevant.
This is the reason why in this paper we consider only tame curves.

\subsection{Some examples}
\label{4.2}

Before we continue the general discussion of plane curves, let us
present some concrete examples of cases which are not unfamiliar in
the physical literature.

We would like first to describe in detail how the genus zero(tree
level) string interactions can be reproduced with a suitable form of
the coefficients in the spectral equation~(\ref{spec}) or~(\ref{pol}).
In the genus zero sector any Riemann surface $\Sigma$ is a punctured
sphere, realized as a $N$-fold branched cover of the $z$-sphere.

\EPSFIGURE{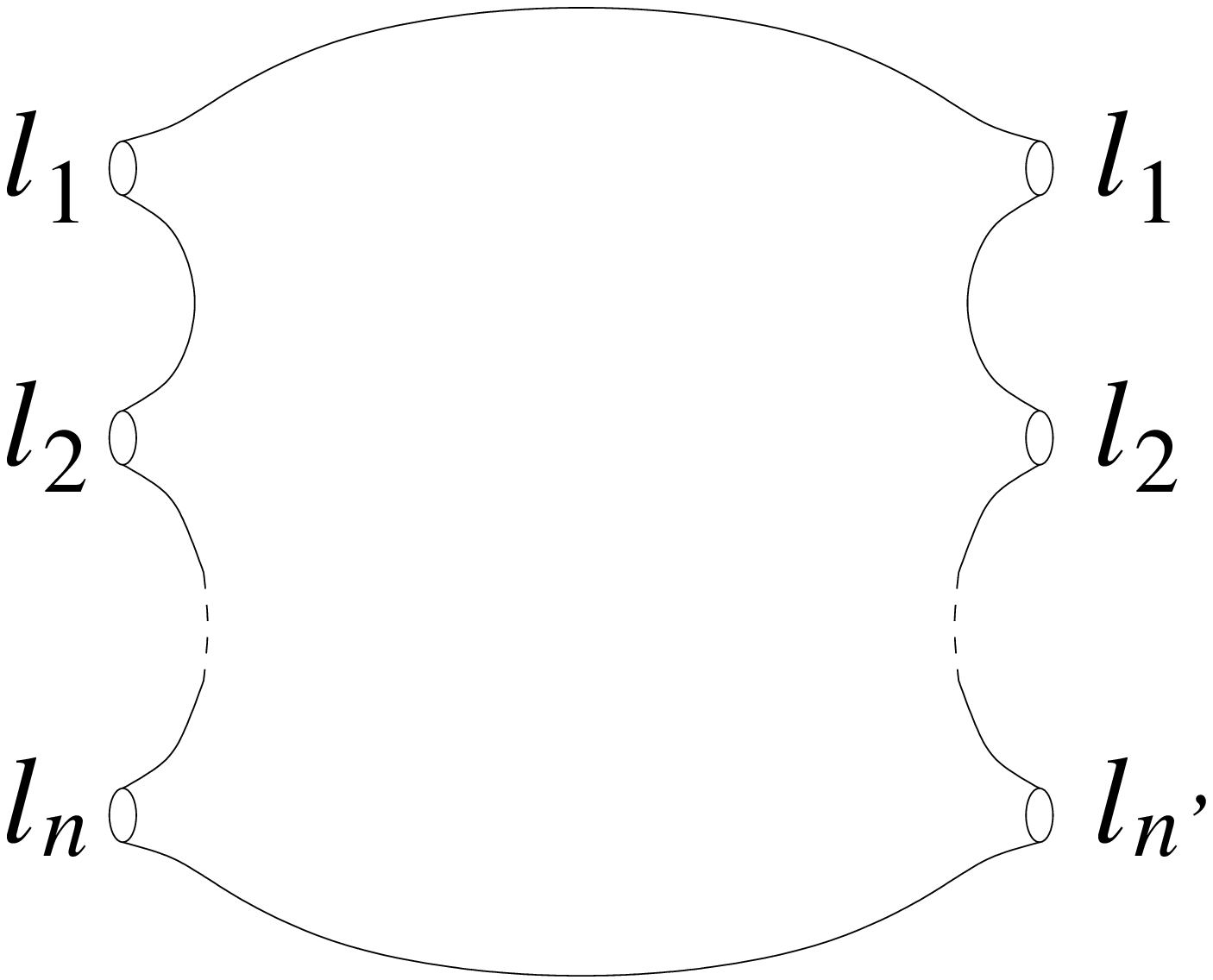,width=11em}{Tree level process of $n$ 
to $n'$ strings.\label{fig1}}

Assume we have $n$ incoming and $n'$ outgoing strings of lengths $l_i$
and $l'_j$, ($i=1\ldots n$, $j=1\ldots n'$), respectively (see 
the figure). From a
phyiscal point of view, we have seen that the length of a string is
interpreted as the + component of its light cone momentum. We recall
that the relation
\be \sum_i l_i=\sum_j l'_j=N\,,
\label{conservation}
\ee
must hold due to conservation of the momentum. We have also seen that
the length of an incoming string $i$ being $l_i>1$ means that the
cover has a branch of order $l_i-1$ at $z=0$, and likewise for
outgoing strings at $z=\infty$.

Our aim here is to construct a polynomial $P$ which underlies such a
string process. Let us tackle this problem by studying the $N$-fold
cover as a holomorphic projection from $\Sigma$ to ${\CP^1}$. As we
have already noticed, the coordinate $z$ does represent such a
projection as a meromorphic function on $\Sigma$: punctures manifest
themselves as zeroes or poles of appropriate orders $l_i$, $l'_j$.
The condition~(\ref{conservation}) means in this picture that the
number of zeroes minus the number of poles, with multiplicity, is zero
(this is the degree of the divisor).

Proceeding in this direction, we construct the generic meromorphic
function in terms of a global coordinate on $\Sigma$, which we can
take to be $y$ itself. This is a useful simplification, which is not
possible in higher genus cases.

The generic meromorphic function satisfying the above requirements on
zeroes and poles is given by the following rational map:
\be 
z = K{(y-y_1)^{l_1}(y-y_2)^{l_2}\cdots(y-y_n)^{l_n}\over
(y-y'_1)^{l'_1}(y-y'_2)^{l'_2}\cdots(y-y'_{n'})^{l'_{n'}}}\,.
\label{merom0}
\ee
This map depends on $n+n'$ parameters, in addition to the constant
$K$: it fixes the $n+n'$ punctures on $\Sigma$ to be located at the
points $y_i$ and $y'_j$. The case of $y_i$ or $y'_j=\infty$ is a
limiting case of the above formula when the relevant factor is absent.
Let us verify that~(\ref{merom0}) gives the right behaviour at $z=0$
and $z=\infty$, see \cite{bbn1}. An example will suffice. Near $y_1$
we can write $z\sim (y- y_1)^{l_1}$, therefore $y\sim y_1 + z^{1/l_1}$,
which is exactly the behaviour considered above.

Now we can make a first exercise of moduli counting. Let us recall
that the moduli space of the Riemann sphere with $p$ punctures is
$p-3$.  To count the moduli in~(\ref{merom0}), we first notice that we
have $n+n'+1$ free parameters. Of these, $K$ corresponds to a
rescaling of the $z$ coordinate; then we can use $PSL(2,\C)$ to
reabsorb three parameters among the $y_i$, $y'_j$. As a result the
meromorphic function describes spheres with $n+n'-3$ moduli, as expected.

\medskip

Now, in order to see whether these curves are reproduced within MST,
we try to cast~(\ref{merom0}) in the form~(\ref{spectr}). One sees
immediately that~(\ref{spectr}) corresponds to curves where one of
the outgoing punctures is at infinity, say $y'_1=\infty$. Given that,
the above map is indeed of the form of~(\ref{spectr}) with
coefficients $a_i$ which are at most linear in $z$:
\be
y^N + a_{N-1} y^{N-1} + \cdots + a_0=0\,,\qquad
a_i = \alpha_iz+\beta_i\,.
\label{spectr0}
\ee

The generic polynomial of this form corresponds to a curve which has
all $l,l'=1$, i.e.\ it has no branches at the $2N$ punctures, and
depends on $2N$ parameters.  Of these, three can be ignored, since
they correspond to transformations that leave $y'_1=\infty$: a
rescaling of $z$; a shift of $y$ and a rescaling of $y$. 
They are the remnant of $PSL(2,\C)$ which keeps $y'_1=\infty$. 

Therefore~(\ref{spectr}), or~(\ref{pol}), contains the right $2N-3$
moduli of spheres.

The cases when some punctures are branched, are limiting cases of the
previous curve when two or more punctures coincide.  This can be
easily seen from the meromorphic map~(\ref{merom0}). Therefore, for
each $l_i>1$, we have to enforce $l_i-1$ conditions on the parameters
$\alpha_i, \beta_i$ of the spectral equation. Thus the free parameters
are, as expected:
\be
2N-\sum_{i=1}^n (l_i-1)-\sum_{j=1}^{n'}(l'_j-1)-3=n+n'-3\,.\0  
\ee
We conclude that at genus zero MST reproduces, via~(\ref{pol}), the
full $n+n'-3$ moduli. 

\medskip

In the case of curves with non-vanishing genus, one would be tempted
to proceed in the same way, that is to construct the meromorphic
projection $\Sigma\to{\CP^1}$ and then invert it. It is rather easy
to construct the meromorphic function $z$ at genus 1. However we come
immediately across a novel feature which was absent in genus 0, but
has dramatic consequences for the moduli counting.

The point is that the punctures on $\Sigma$, represented as zeroes and
poles of the meromorphic function, cannot be arbitrary. This is a
feature of the torus and of higher genus curves. There is a condition
that they have to satisfy, which is the price we have to pay to be
able to represent the punctured surface as an algebraic
curve. Mathematically speaking, the divisor of a meromorphic function,
is not a generic divisor of degree zero, but is a principal one, which
amounts to some extra condition on the punctures.  The same condition
was absent on the Riemann sphere because there every divisor of degree
zero is principal.

To see which condition appears, let us represent explicitly the
meromorphic function using a coordinate $t$ taking values in the
fundamental parallelogram. On a torus a meromorphic function can be
written as the ratio of products of ``translated'' theta functions:
\be
z = K{\t(t-t_1)^{l_1} \t(t-t_2)^{l_2} \cdots \t(t-t_n)^{l_n}\over
    \t(t-t'_1)^{l_1} \t(t-t'_2)^{l_2} \cdots \t(t-t'_{n'})^{l'_{n'}} }\,,
\label{merom1}
\ee
(see the end of section~\ref{51} for details). Now, for $z$ to be
single valued, the $t_i$, $t'_j$ have to satisfy a condition, that is
the vanishing of the Abel-Jacobi map: 
\be
\sum_i l_i t_i -\sum_j l'_j t_j = 0 ~~ \hbox{mod} ~~ \Gamma\,.\label{g1}
\ee
where $\Gamma$ is the group of periods, which for the torus is the 
usual lattice of complex translations: $\Sigma={\mathbb C}/\Gamma$.

\EPSFIGURE{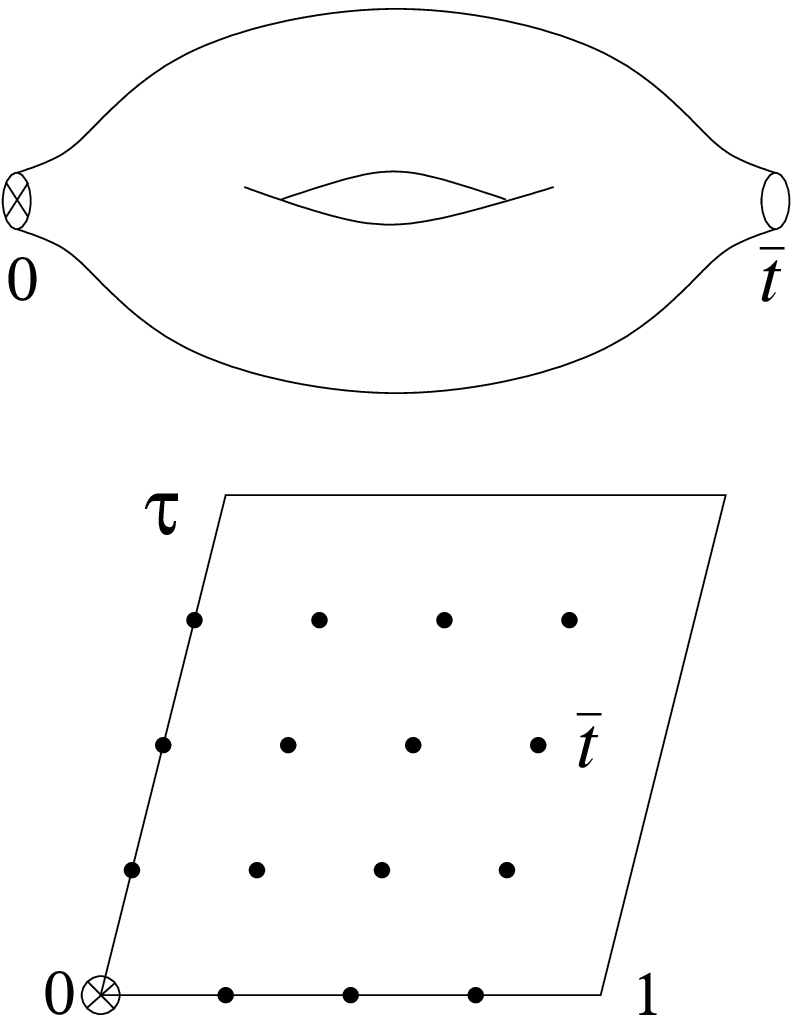,width=11em}{Discrete modulus}

It is instructive to look at the case of the propagator of a long
string at genus one. We require the insertion on the torus of an
incoming and an outcoming string of length $l$, at two points. By
translation we can bring one of them at the origin and the other at,
say, $\bar t$.  The above condition is in this case:
\be
l\bar t=0 ~~\hbox{mod}~~ \Gamma\,,
\ee
and we see that $\bar t$ has to lie on the lattice $\Gamma/l$
indicated in the picture.

We can see from here that at finite $N$ we have some limitations on the
possible diagrams we can realize, however as $l$ and $N$ become large,
the lattice $\Gamma/l$ fills the plane and we recover the continuous
modulus. 

In section~\ref{5} we will discuss in general the limitations of
this kind. Therefore we leave this subject at this point and discuss
other aspects concerning genus one curves.

The next thing we would like to do is to mimic the genus 0 case by inverting 
eq.~(\ref{merom1}). This is certainly possible locally, but, unlike the genus 0
case, we will not find in general a polynomial equation of the type~(\ref{spec}).
Therefore constructing the meromorphic projection~(\ref{merom1}) gives us only
limited information about plane curves. In fact, what one expects is 
that the plane curve corresponding to~(\ref{merom1})
is in general in a singular representation (see below). 

It is then necessary to study singular plane curves.

\subsection{Plane curves and their representation}
\label{4.3}

At the beginning of this section we have called plane curves the locus
of points which are solution of an equation like~(\ref{pol}) in
${\mathbb C}^2$. This definition is too generic and lends itself to
ambiguities. For example, we know the coordinates $y$ and $z$ are not
on the same footing in MST.  A $z$ rescaling (at strong coupling) is a
symmetry of any process in MST, but no other $pSL(2, {\mathbb C})$
transformation is a symmetry transformation of a string process ($z\to
1/z$ is a symmetry transformation of the theory, not of a single
process). As for $y$ it is not clear which coordinate transformations
are a symmetry.\footnote{For instance, a $y$ inversion is not allowed,
since it generates poles violating the holomorphicity the $a_i$'s.}

We resolve this and other ambiguities by embedding our curves in
${\CP^2}$: we introduce the homogeneous coordinates $x_0,x_1,x_2$ with
$z=x_1/x_0, y=x_2/x_0$. By multiplying~(\ref{spec}) by a suitable
power of $x_0$ we obtain the equation of the curve in $\CP^2$ in the
form
\be 
F(x_0,x_1,x_2)=0\,. \label{pol1}
\ee

Then the coordinate transformations that do not change the curve are
in general those of $PGL(3,{\mathbb C})$. However, as we said above,
the points $z=0, \infty$ should be fixed in MST. This means that
$x_0=0$ and $x_1=0$ should not be modified by any transformation. In
conclusion the coordinate transformations that give rise to physically
indistinguishable processes in MST, are those of the subgroup ${\cal
H}\subset PGL(3,{\mathbb C})$ defined by
\be
\left( \matrix{x_0'\cr x_1'\cr x_2'\cr}\right )= 
\left(\matrix{* & 0& 0\cr 0&*&0\cr *&*&*\cr}\right)
\left( \matrix{x_0\cr x_1\cr x_2\cr}\right ).\label{lintransf}
\ee
In terms of $y$ and $z$, these transformations include 
rescalings of $y$ and $z$ and linear transformations
$y\to y +\alpha z +\beta$, with complex constants $\alpha$ and $\beta$. 
They are acceptable coordinate transformations which involve 4 complex 
parameters. This fits in our counting of the independent parameters in the
previous subsection.
 
From now on, although we keep speaking mostly in terms of $y$ and $z$,
we always understand the corresponding formulation in terms of
$x_0,x_1,x_2$.  For example, a transformation like $z\to 1/z$ must be
accompanied by $y\to y/z$ in order for us to remain within
${\CP^2}$. The latter is a compact space, therefore embedding the
curves in it means compactifying them by filling the punctures with
suitable points in ${\CP^2}$.  

Given a curve defined by~(\ref{pol1}), the points in it where all
partial first order derivatives vanish are {\it singular points}. See
Appendix for a short summary on singularities. Singularities will play
an important role in the following. For example, eq.~(\ref{merom1})
above, when written in homogeneous coordinates reveals a singularity
corresponding to the point $z=\infty$.

The information about the branch points of a curve is contained in the
discriminant. The discriminant $\delta$ of~(\ref{spec}) is
proportional to $\Delta^2$, where $\Delta$ was defined in
(\ref{Delta}). For a definition of the discriminant in terms of the
$a_i$'s, see for example \cite{gel}.  The zeroes of the discriminant
define the branch points and their multiplicity gives the multiplicity
of the branch points.

A useful tool in studying plane curves is the {\it Newton Polygon}.
Let us consider the polynomial $P(y,z)$ in~(\ref{pol}). We associate
to each monomial $z^\alpha y^\beta$ in it a point $p=\alpha$,
$q=\beta$ in a ${p,q}$ plane. We obtain a set of points called the
{\it carrier}: its convex hull is by definition the {\it Newton
polygon} associated to the curve. From the Newton polygon one can
deduce a lot of information concerning the curve. For the curves we
consider the Newton polygon always contains the point $(p=0,q=N)$ and
is contained in the equilateral triangle formed by the $p$ and
$q$-axis and by the line $p+q=N$.

\FIGURE[r]{\epsfig{file=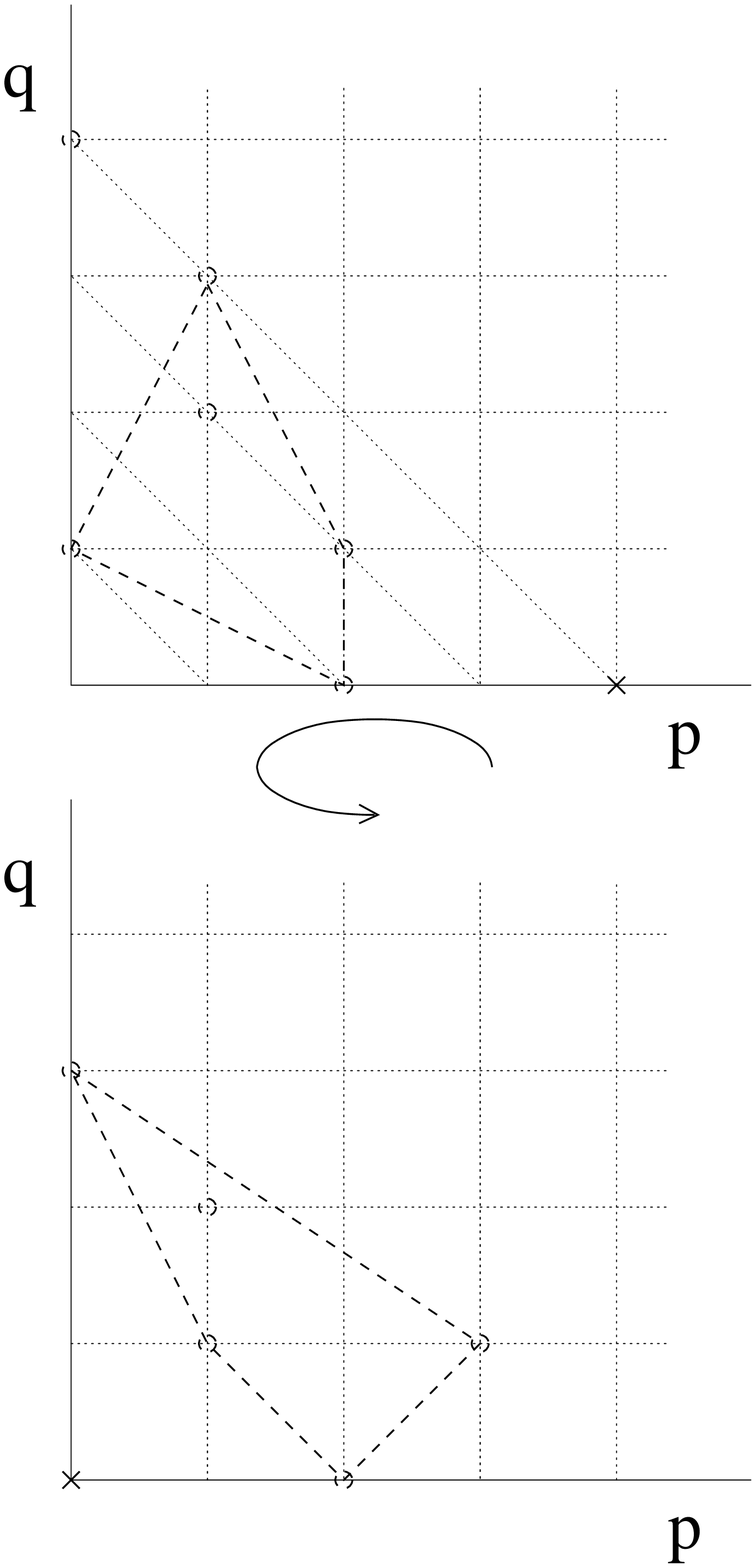,width=10em}
\caption{Newton diagrams and transformation of coordinates.}
\label{trans}}

If the curve is irreducible (we will always consider only irreducible
curves), the Newton polygon must contain at least one point in the
$p$-axis. Therefore for standard curves the Newton polygon will be
formed by an upper broken line and a lower broken line (the latter by
definition may contain segments of the $q$ and $p$-axis).  For
example, if the point $(p=0,q=0)$ is not in the carrier it means that
the point $y=z=0$ belongs to the curve. If, in addition, one of the
points $(p=1,q=0)$ or $(p=0,q=1)$ does not belong to the carrier, the
point $y=z=0$ is obviously singular. If it is so, the singularity is locally
given by the product of the local curves corresponding to the various
sides of the lower line. In addition to that, each side can contain a number $m$ of
points of the integer lattice, beside the vertices, which may or may
not be present in the carrier. The number of the components
corresponding to this side is simply $m+1$ .

If one wants to control what happens over ${z=\infty}$ in a given curve, 
one has to find what the intersections with the line  $x_0=0$ are; the
polynomial which describes such intersection is simply given by the points of 
the carrier which lie on the upper line of the Newton polygon.
It is useful to see what the Newton diagram looks like in new affine 
coordinates around the point. This is easily done by means of transition
functions: $z^\prime = 1/z$, $y^\prime = y/z$. Figure \ref{trans} shows 
an example: diagonal lines in the first diagram become the
new vertical lines in the second one (one has to multiply by $z^N$ in order to
recover a polynomial).
 
In addition to that, suppose for instance
that the point ${p=N}$, $q=0$ does not belong to the carrier; the Newton polygon
around the point ${x_0=0, x_2=0}$, which now is in the curve, after going to
the affine chart in which $x_1=1$, is simply given
by a linear deformation and a reflection around the line $p=N$. In particular,
in order for this point not to be singular, one of the points ${p=N-1,q=0}$ 
and ${p=N-1,q=1}$ has to be in the carrier.

After this generic information about Newton polygons, let us see some concrete
examples of genus 1 processes.

\subsection{Examples: smooth elliptic curves}  
\label{4.4}

We start with the case 
${N=3, g=1}$, for which there is already a good variety of examples.
These have the advantage that one can check the results by explicitly solving 
the cubic algebraic equation by means of Cardano's formula.
We do not write down the algebraic equations, but simply the corresponding 
polygons. The coefficient of the monomials within or on the border of the
Newton polygon are understood to be generic,
unless otherwise specified.

\FIGURE{\epsfig{file=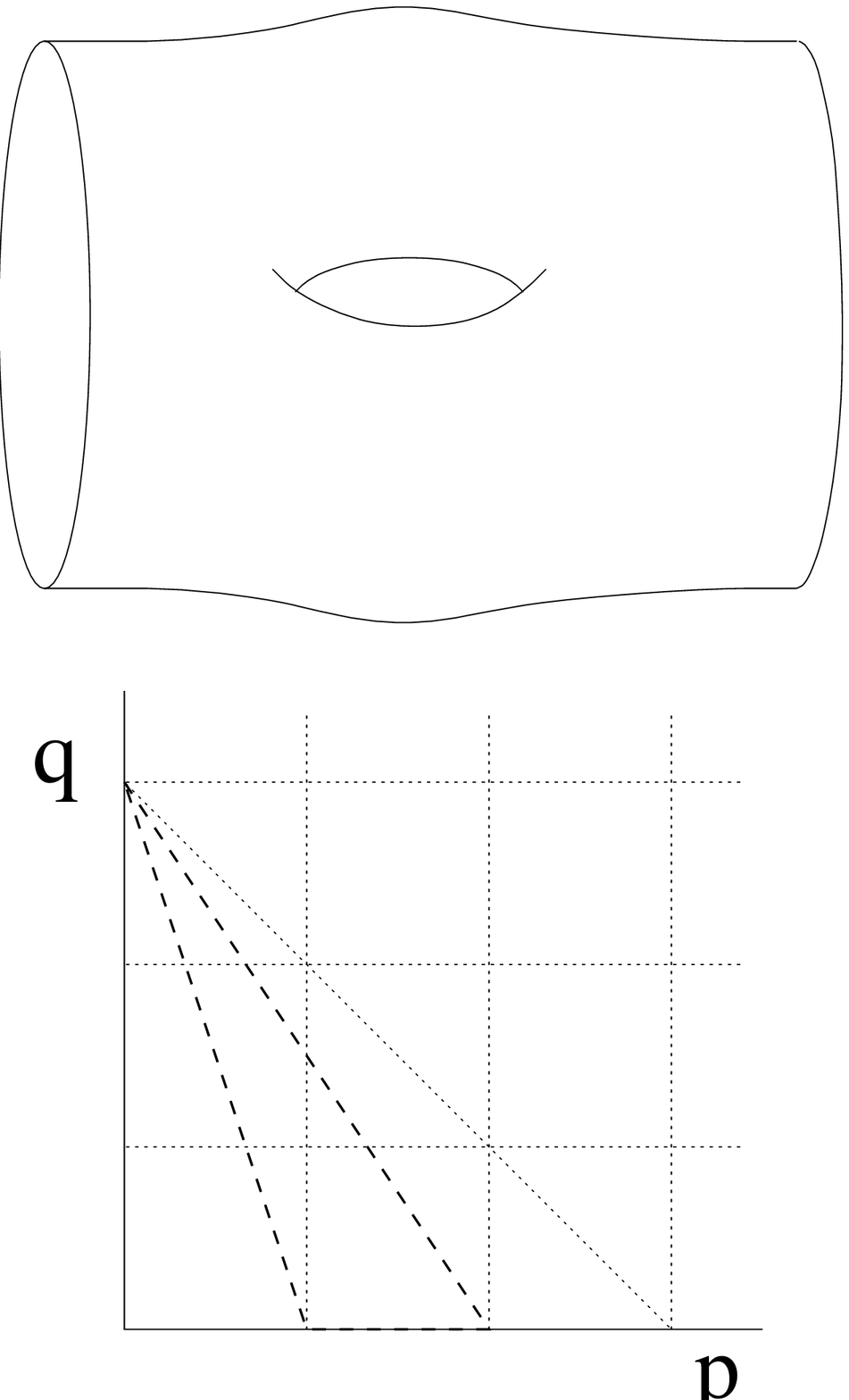,width=9.5em}
\caption{Self-energy of a string.}
\label{total}}

The simplest process one can imagine is the string self-energy. This 
means that we have to look for a totally branched curve over $z=0$ and 
$z=\infty$. Remember that the polynomials giving the solutions over these
points are given by the points of the carrier on the $q=0$ and $p+q=N$ lines
respectively. So one simple solution is given by the carrier shown in figure
\ref{total}; the generic case 
will be non-singular also at finite $z$ and so the genus will be one.
The presence of the points $(1,0)$ and $(2,0)$ ensures the nonsingularity of
$0$ and $\infty$; the local behaviour around them is given by the upper side
of the inner (shaded) triangle.

Next we would like to describe a joining of strings. In this case we keep
$z=\infty$ totally branched, while we add a point on the $q$ axis in order to
have, at $0$, a polynomial like $y^3 +y^2$ instead of $y^3$, so that $y=0$
appears twice as a solution and $y=-1$ once. 
It is now easy to construct all combinations: figure \ref{various} shows
various examples and their Newton polygons --- as above, with generic
coefficients.

\FIGURE{%
\epsfig{file=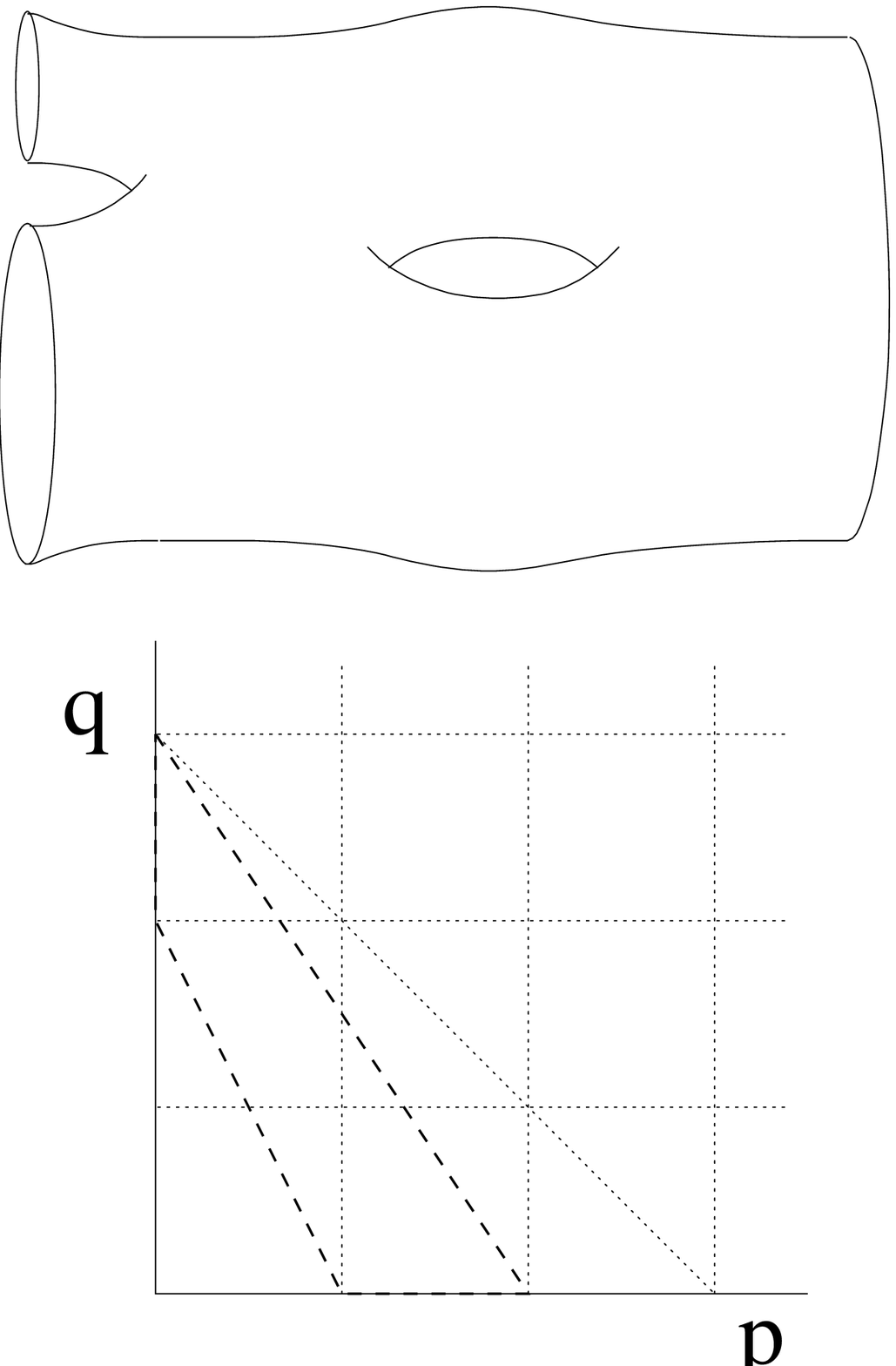,height=15em}\hspace{1em}
\epsfig{file=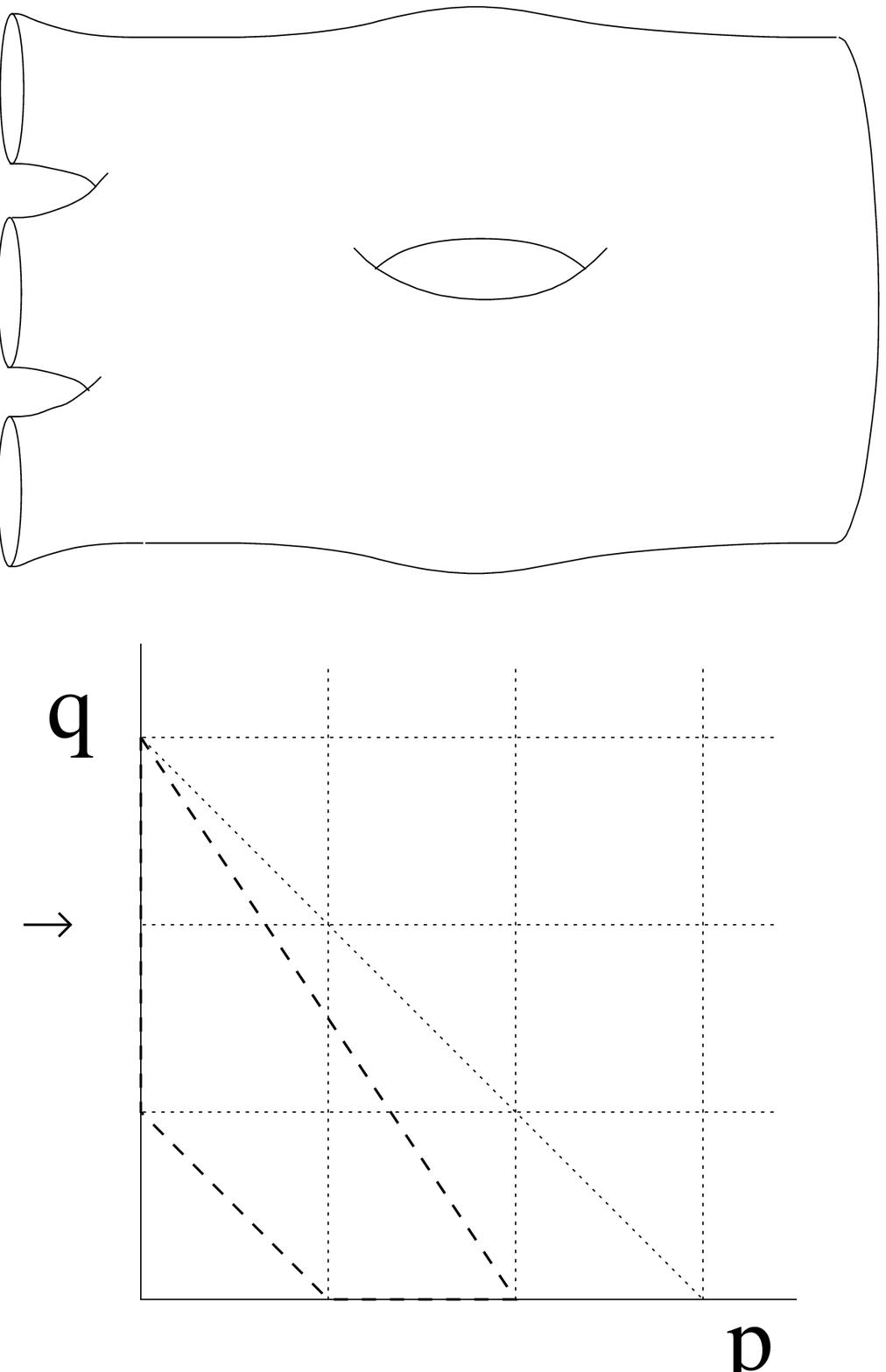,height=15em}\hspace{1em}
\epsfig{file=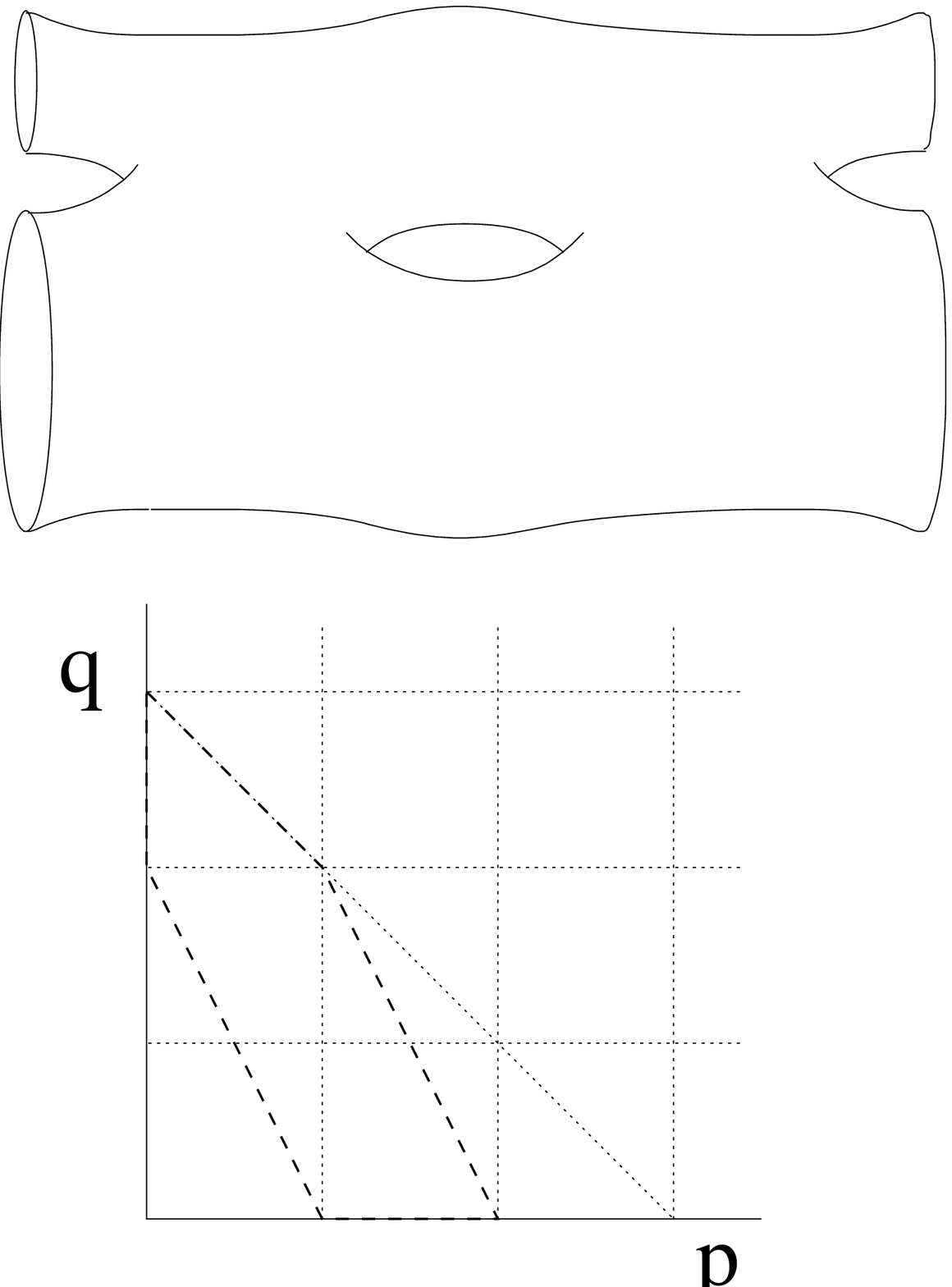,height=15em}
\caption{Examples of processes and their Newton diagrams.} 
\label{various}}

\subsection{Examples: singular plane curves}
\label{4.5}

Of course for some choice of the parameters a singularity may appear.
In this case one has simply to replace the finite hole in these
figures by a hole shrunk to a point (for example see
fig.~\ref{quartic}); the curve becomes genus zero, i.e. a sphere with
two identified points. This singularity is the simplest one, it is
characterized by a non-vanishing Hessian and is called a {\it node}.
All nodes can be viewed as two points identified: blowing up a node
amounts to separating the points.  For instance, consider our first
case (figure \ref{total}): the polynomial which corresponds to the
diagram can be written as
\begin{equation}
P= y^3 + c z y + z (z-a)\,,.
\end{equation}
Imposing that a point be singular, one finds that a necessary (but not sufficient)
condition is that its discriminant, $\delta= z^2 [27(z-a)^2 - 4 c^3 z]$, has
a multiple root. The double root at $z=0$ just signals that this point is 
another 
triple branch point, as we already knew; imposing that the remaining factor be
a square, one finds several values, of which for instance $c=0$ gives a triple
branch at $z=a$ and no singular point, and $c= -3 a^{1/3}$ gives
instead a node.

\FIGURE{\epsfig{file=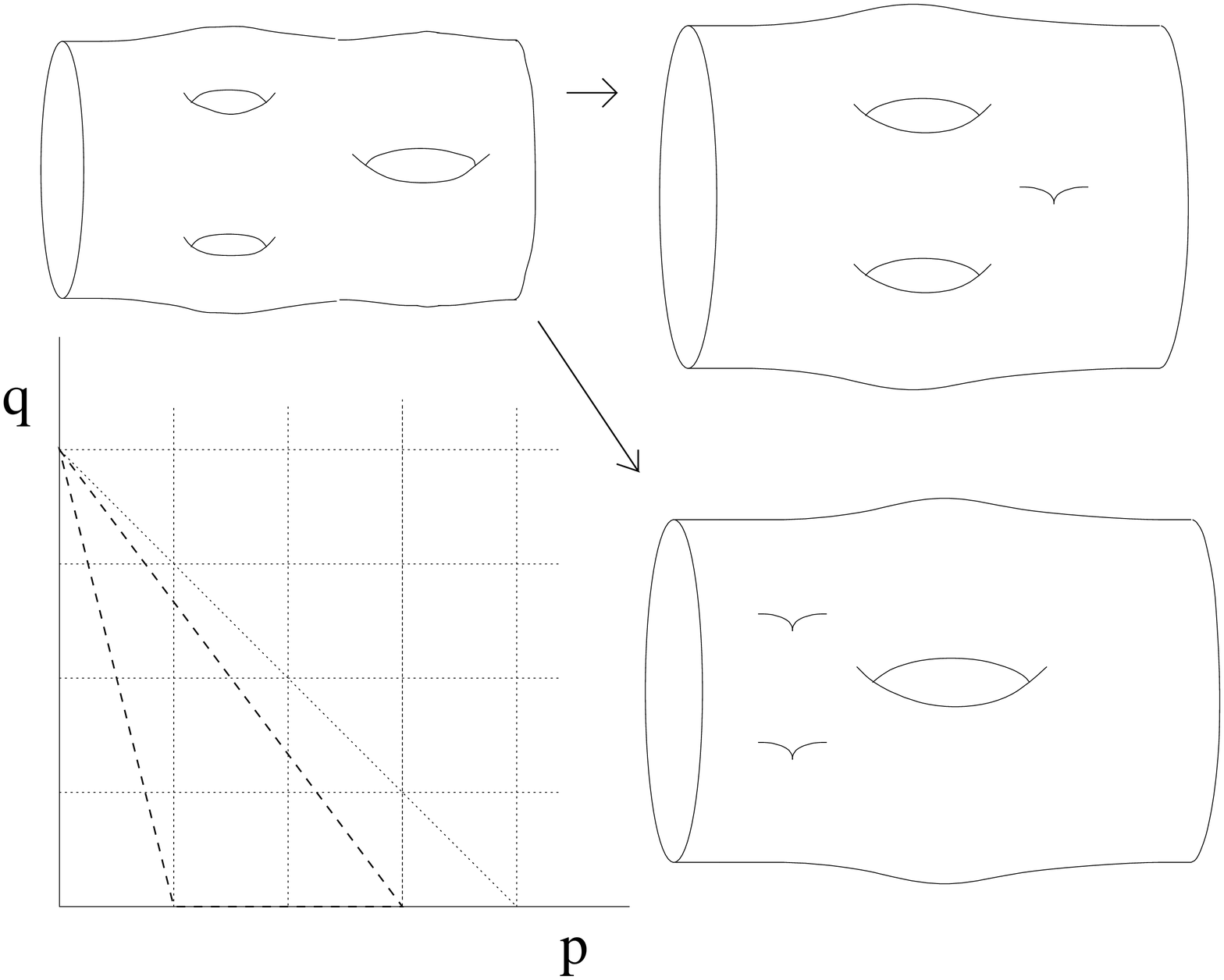,width=17em}
\caption{Shrinking cycles: totally branched quartics at genus two and one.}
\label{quartic}}

This introduces us to our next task: to show how it is possible to
describe low-genus highly branched curves. We will describe in detail
the self-energy case. We take $N=4$; since we want total branching we
can choose a diagram like that in figure \ref{quartic}. The
corresponding polynomial has the coefficients corresponding to the
vertices of the polygon, and can also have coefficients corresponding
to the points on the sides or in the interior (by the way the latter
are always $(N-1)(N-2)/2$ in number if there is no singular point at
$0$ and at $\infty$ and count the genus of the corresponding smooth
curve).  Now we can look for singular cases in this family along the
lines of the previous example; since already in this case computations
become complicated, we restrict ourselves to the biquadratic case. In
other words the polynomial we start with is
\begin{equation}
P= y^4 + b z y^2 + z(d+ e z + f z^2)\,;\0
\end{equation}
its discriminant is 
\begin{equation}
\delta= 16 z^3 (d+ e z + f z^2)(4 d + 4 e z + 4 f z^2 - b^2 z)^2\,.\label{discr}
\end{equation}

As before, the term $z^3$ shows that the branching at $z=0$ is of
order three, i.e.\ four sheets meet there.  The other two terms mean
the following. Solutions of a biquadratic equation are in general $\pm
y_{1,2}$. Its discriminant can vanish in two cases: if $y_1=y_2$ or
$y_1=-y_2$ --- this is determined by the third term in~(\ref{discr})
--- in which case, at the corresponding value of $z$, there is a
couple of double branch points; if $y_1=0$ or $y_2=0$, which is
determined by the second term, there is a single node. If we choose
the coefficients so that the third term is a fourth power, we have two
nodes, and so genus one; if, instead, the coefficients are chosen so
that the second term is a square, we have a single node, and so genus
two. The situation is shown in figure~\ref{quartic}.

If one does not wish to restrict to this particular case, one can
still find examples of genus 1 and 2 curves. One notes, for instance,
that imposing a node first and a total branch in $z=0$ afterwards is
computationally easier. A totally branched, non biquadratic genus one
quartic is for instance given by
\begin{equation}
P = y^4 -5 z y^2 + 3 z(z+1) y - \frac{z}{2} (z+1)^2 \,,
\end{equation}
which has two nodes at $\{z=-1,\, y=0\}$ and $\{z=1,\, y=1\}$, and two
regular branch points: $\delta=z^3(z+1)^2(z-1)^2(63z^2+62z+63)$.

\subsection{The role of singularities}

We believe the above examples are sufficient to illustrate 
the problems connected with the representation of Riemann surfaces
with punctures by means of plane curves. It is impossible in general to
represent Riemann surfaces by means of smooth plane curves embedded in
the two complex dimensional space spanned by the coordinates $y$ and $z$.
One can say that singular plane curves within stringy instantons are 
the ordinary tools MST uses in order to reproduce the string interaction 
configurations required by string theory (actually, as will be seen in 
the next section, only in the $N\to \infty$ limit is this completely true).
 
Far from representing a problem, singular plane curves are most
welcome.  They come with a gratifying bonus: the solution of a serious
problem for the identification of MST at strong coupling with string
theory.  This identification is possible if string theory is
formulated in the light-cone gauge. In MST,~(\ref{eSYM}), ten
dimensions enter into the game, two world-sheet dimensions plus eight
transverse dimensions represented by the (diagonal) $X^i$. At first
sight they seem to have a different nature, however it is clear that
in a light-cone framework the two world-sheet dimensions are to be
interpreted as representatives of the time and longitudinal
dimensions, denoted 0 and 9, which bring the total of physical
dimensions to ten. Now, stringy instantons characterized by a smooth
plane curve, extend over four out of these ten dimensions. In other
words it would seem that MST at strong coupling can only describe
four-dimensional string processes.  If this were true it would be
hard to justify it in the light of the correspondence MST --- string
theory.

However here come singular curves to our rescue.  
Singular curves become smooth if one enlarges the space where they are 
embedded. The standard way to resolve a singularity is to blow it up 
(see Appendix), which means that a singular point is replaced by
a two-dimensional sphere. For example, we have already pointed
out that curves in ${\CP^2}$ with nodes (a node is the simplest 
possible type of singularity) only, can be smoothed out by embedding them 
in ${\CP^3}$, i.e. by adding two dimensions. It is natural to interpret 
this by saying that the corresponding string process extend over six 
(instead of four) dimensions. It is not difficult to imagine processes 
that extend over more (up to ten) dimensions. To better convince
ourselves of this fact we can take the reverse point of view. 
Suppose we want to embed these higher (than four) dimensional 
processes within the instantons of the 2d field 
theory~(\ref{eSYM}). The only possibility is to squeeze (project) them to
the appropriate four dimensions: such operation of projecting gives 
rise to singularities. It goes without saying that the true significance 
of singular plane curves is given by their representing higher (than four)
dimensional processes.
 
This is in particular true for punctures. Singularities that occur in 
the counterimage of $z=0,\infty$ represent overlapping punctures, and in 
order to disentangle them one has to enlarge the embedding space. 
It is clear what happens: a high dimensional string process squeezed to
four dimensions may require that the locations of incoming and outgoing 
strings overlap.

\section{The moduli space of MST}
\label{5}

We have seen in section~\ref{2} and~\ref{4} that for finite $N$ the
genus of the plane curves that appear in MST has an upper bound given
by ${1\over 2} (N-1)(N-2)$.  We also anticipated in section~\ref{2}
that the moduli space of plane curves of genus $h$ with $n$ punctures
turns out to be a discretized version of the moduli space of Riemann
surfaces with the same topological type. In this section we want to
examine this point. In the first part we see the origin of the
discretization, in the second part we confirm this result by
estimating the dimension of the moduli space of stringy instantons on
the cylinder.

\subsection{Discretization}
\label{51}
 
A very convenient way to proceed is to make a comparison with the
Mandelstam param\-etrization of the moduli space of Riemann surfaces
with punctures \cite{mand,dhofo,GSW}.  To this end, let us first
review some basic facts about the realization of Mandelstam
diagrams. We refer to \cite{gidwol} for a complete account of the
following very quick review, after which, we will examine the
consequence of the main new input from MST, that is holomorphicity of
the covering map which defines the Mandelstam diagram.  The result
will be a set of constraints on the kinematical data of the diagram
which turn out to be a quantization condition for some of the
Mandelstam parameters. In the large $N$ limit these constraints loosen
their effectiveness and allow us to recover the full moduli space of
the string diagrams.

Let $\Sigma$ be a compact Riemann surface of genus $h$ and let
$\omega_I$, $I=1,\dots,h$ be a set of holomorphic differentials on
$\Sigma$ normalized by $\oint_{\alpha_J}\omega_I=\delta_{IJ}$, while
$\oint_{\beta_J}\omega_I=\Omega_{IJ}$ is the period matrix.  We fix
$n$ punctures $\{ Q_1,\dots,Q_n \}$ on $\Sigma$ and define the divisor
$D=Q_1\cdot\dots\cdot Q_n $. We also introduce a set of $n$ real
numbers $R=\{r_1,\dots,r_n\}$ such that $\sum_i r_i=0$.

Now, let $\omega$ be the differential which is holomorphic on
$\Sigma\setminus D$ with simple poles at $D$ with ${\rm
res}_{Q_i}\omega=r_i$ and ${\rm Re}\oint_{\alpha_I}\omega=0={\rm
Re}\oint_{\beta_I}\omega$.

In \cite{gidwol} it was shown how the above differential defines a
nice procedure which allows us to look at $\Sigma$ as a {\it
topological} covering of a cylinder: one can easily decompose $\Sigma$
into pants along the level lines of the function $\tau(P)\equiv{\rm
Re}\int^P\omega$. In this sense, $\omega$ induces on $\Sigma$ the
structure of a Mandelstam diagram.  The Mandelstam parameters are the
twist-angles $\theta_b$, $b=1,\ldots 3h +n-3$, along the junctures of
the pants decomposition and the relative `time' coordinates $\tau_a-\tau_0$,
$a=1,\ldots, 2h+n-3$, of the $2h+n-2$ interaction points. $h$
additional real parameters are the internal light-cone momenta
$p_I^+=\oint_{\alpha_I}\omega$. Altogether they form a set of
$6h-6+2n$ real parameters. In \cite{gidwol} it was shown that these
parameters represent good coordinates on the moduli space of genus $h$
Riemann surfaces with $n$ punctures ${\cal M}_{h,n}$.

To complete the picture we identify the set $R$ with the + components
of the external light-cone momenta of the diagram, i.e. the periods of
$\omega$ around the punctures. We also have the relations
$\oint_{\beta_I}\omega={i\over 2\pi} p^+_K {\cal W}^I_{Kb}\theta_b$,
where ${\cal W}^I$ are integer-valued matrices which depends on the
pants decomposition of the Riemann surface and its intersections with
the $\alpha$ and $\beta$ cycles.
 
Our strategy now is the following. We first construct an explicit form
for $\omega$, in terms of the prime-form, the $\omega_I$'s and the
period matrix of $\Sigma$. Then we compare this $\omega$ with the one
that comes from MST. The relevant new input consists in the fact that
MST induces on $\Sigma$ the structure of a {\it holomorphic} covering
of the Riemann sphere (as usual we consider the latter instead of the
cylinder). By this we mean that, if $z:\Sigma\,\to\CP^1$ is the
covering map in the MST scheme, the coordinate $z$ is a meromorphic
function on $\Sigma$. The role of $\omega$ in MST is played by $d{\rm
ln}z$, therefore we have to identify them.  This condition becomes a
constraint on the data of the Mandelstam diagram. In fact, it means
that $D^R\equiv Q_1^{r_1}\cdot\dots\cdot Q_n^{r_n}$, being the divisor
of the meromorphic function $z$, is a principal divisor on $\Sigma$, so that
in particular $r_i\in {\mathbb Z}$. As a consequence, some constraints
appear in the data of the Mandelstam diagram and these conditions
induce a complex codimension $h$ slicing of the moduli space. This can
be seen as follows.

Let $\omega_{P_+P_-}$ be the holomorphic differential on
$\Sigma\setminus \{P_+,P_-\}$ with simple poles at $P_\pm$ with
residues $\pm1$ and imaginary periods. It can be written
as\footnote{$H(P,P_+,P_-)$ depends also on a base point which is
irrelevant in this context and, for the sake of simplicity, is not
specified.}
\be
\omega_{P_+P_-}(P)=d_{(P)} {\rm ln}
\left[
{E(P,P_+)\over E(P,P_-)}\cdot \e^{
2\pi i {\rm Im}\int_{P_-}^{P_+}\omega_I {\Omega^{(2)}}^{-1}_{IJ}
\int^P\omega_J
}
\right]
=d_{(P)}{\rm ln} H(P,P_+,P_-)
\, ,
\label{opppm}
\ee
where $E(P,Q)$ is the prime form on $\Sigma$, $\Omega^{(2)}$
is the imaginary part of the period matrix and $d_{(P)}=dP 
\cdot{\partial\over \partial P}$.

In terms of the above differentials we can write  
\be
\omega=\sum_{l=1}^{n-1} k_l \omega_{Q_lQ_{l+1}}\, ,
\label{oopp}\ee 
where $k_i-k_{i-1}=r_i$ and $k_0=0=k_n$;
substituting~(\ref{opppm}) into~(\ref{oopp}) we obtain
$$\omega(P)= d_{(P)} {\rm ln} \tilde z(P)$$ 
where
\be
\tilde z(P) = \prod_{l=1}^{n-1}
\left[ H\left(P,Q_l,Q_{l-1}\right)\right]^{k_l}.
\label{ztilde}\ee

Now, as anticipated above, we make the identification $\omega = d \ln z$.
This requires that $\tilde z = z$ up to a multiplicative constant, which implies
that $\tilde z$ is a well defined meromorphic function on $\Sigma$.
On the one hand this imposes that the residues $r_l$ be quantized in integer 
values. On the other hand it requires that the differential
$d\tilde z$ have vanishing periods along $\alpha$ and $\beta$ cycles.
The latter condition is fulfilled iff 
\be
\sum_{l=1}^n r_l\int^{Q_l}\omega_I=m_I+n_J\Omega_{JI}
\label{prdi}\ee
for some $m_I,\, n_I\in {\bf Z}$. 
At this point the situation is clear:~(\ref{prdi}) is the vanishing 
condition for the Abel map and says that the divisor $D^R$ is principal.

Conversely, let $z$ be a meromorphic function on $\Sigma$ and $D^R$
its divisor. By definition~(\ref{prdi}) holds and ${\rm res}_{Q_l}
d_{(P)}{\rm ln}z= r_l$.

Notice that the periods of $\omega$ are quantized in integral values as
\be
\oint_{\alpha_I}\omega=2\pi i n_I
\quad {\rm and} \quad
\oint_{\beta_I}\omega=-2\pi i m_I\,,
\label{perqua}\ee
and this condition is equivalent to~(\ref{prdi}).

Eq.~(\ref{perqua}) means that the internal light-cone momenta of the 
diagram are quantized and that, in addition, there are $h$ discretizing 
constraints on the twist-angles of the Mandelstam diagram.
Since these variables, together with the relative interaction times which 
have been left untouched, are the coordinates of the moduli space,
we are left with a discrete slicing of the moduli space
${\cal M}_{h,n}$, each slice being of complex dimension $2g-3+n$. This
discretized moduli space is what we have called ${\cal M}^{(h,n)}_{N}$
in section~\ref{2}.

In the large $N$ limit, however, the quantization condition disappears
in a continuum of values\footnote{In some sense, the topological
reconstruction of the Mandelstam diagram can be seen as an
infinitely-sheeted holomorphic covering of the cylinder.}
\be
{\rm lim}_{N\to\infty}{1\over N}\left[ {\bf Z}^h\bigoplus 
\Omega {\bf Z}^h\right]={\bf C}^h\,.
\label{limjac}\ee
Simultaneously, for large $N$ also the
bound ${1\over 2}(N-1)(N-2)$ on the genus of the plane curves in MST, becomes
ineffective, and we recover the full moduli space of string theory. 

It is interesting to review the genus 0 and 1 case in detail.

\paragraph{Genus zero}

On the sphere the prime form is simply
$E(P,Q)=P-Q$ and then, up to a multiplicative constant,
\be
z=\prod_{i=1}^n (P-Q_i)^{r_i}\,.
\label{genus0}
\ee
Splitting the divisor $D^R=D_0\cdot D_\infty^{-1}$ into its zero and
polar parts --- where $D_0=\prod_{r_i>0} Q_i^{r_i}$ and
$D_\infty=\prod_{r_i<0} Q_i^{-r_i}$ are its positive and negative
parts respectively --- we get
\be
z={ \prod_{i|r_i>0} (P-Q_i)^{r_i}\over \prod_{i|r_i<0} (P-Q_i)^{-r_i}}
\label{genus0`}
\ee
and we recover the result~(\ref{merom0}). The counting of section~\ref{4.2} tells us that
the independent complex parameters are $n-3$. Therefore in this case 
there is no moduli quantization at all and the moduli of plane curves
cover the full moduli space ${\cal M}_{0,n}$. 

These curves correspond to tree level Mandelstam diagrams with
$n$ external strings each of light-cone momentum $r_i$.
Therefore, in this case, MST gives exact results at
finite $N$ (except for the fact that the + components of the external 
momenta are discrete).

\paragraph{Genus one.}

The first case in which moduli quantization becomes effective is at $h=1$.
Let us specialize to the torus the above construction.
In this case, the prime form is proportional to the odd theta function
$E(P,Q)\propto\Theta_{odd}(P-Q|\tau)$ and there is a unique holomorphic
differential $\omega_1=dP$. Therefore 
\be
\omega_{P_+P_-}(P)=d_{(P)} {\rm ln}
\left[
{\Theta_{odd}(P-P_+|\tau)\over \Theta_{odd}(P-P_-|\tau)}\cdot {\rm exp}
\left(
2\pi i {\rm Im}\left(P_- - P_+\right) ({\rm Im}\tau)^{-1}(P-P_0)
\right)
\right]
\, ,
\0\ee
and we get
\be
\omega(P)=\sum_{l=1}^{n-1} k_l \omega_{Q_lQ_{l+1}}(P)=
d_{(P)} {\rm ln}
\prod_{l=1}^{n}  
\left[
\Theta_{odd}(P-Q_l|\tau)
\right]^{r_l}
\cdot \e^{
r_l
2\pi i {\rm Im}\left(Q_l\right) ({\rm Im}\tau)^{-1}P
}=d_{(P)} {\rm ln} z\,.
\0\ee 
Using the standard modular properties of $\Theta$-functions
one obtains that $z$ is a well defined meromorphic function
on the torus $T_\tau\equiv{{\bf C}\over <1,\tau>}$
iff $r_l\in{\bf Z}$ and 
\be
\sum_{l=1}^n r_lQ_l=m+n\tau
\label{prdigu}\ee
for some integers $n$ and $m$. This is the discretizing condition for the
the genus one case. It coincides with the condition already found in
section~\ref{4}, eq.~(\ref{g1}). 

One can verify that in all the genus one examples we have considered
in secs.~\ref{4.4} and~\ref{4.5}, the counting of independent
parameters matches the formula $2h+n-3$. We believe that, for any
topological type ($h,n$), one can construct plane curves with $2h+n-3$
independent parameters.

\subsection{The dimension of the moduli space of stringy instantons}

In this section we verify the correctness of the result obtained above 
by counting the dimensions of the solution space of Hitchin equations 
on a cylinder up to gauge transformations in the strong coupling limit. 
The method is based on
zero modes counting, therefore it is sensitive only to continuous
dimensions. Since the group factor $Y$ does not contain free parameters
we expect to find a $2h+n-3$ dimensional space of solutions.  

The strategy is the following: first we linearize the system by 
transferring the calculation to the tangent space, 
then we calculate the dimension of the subspace of the tangent space
which is orthogonal to infinitesimal gauge transformations, 
in the strong coupling limit $g\to\infty$.
The equations in the $g\to\infty$ limit are in a form 
which can be lifted to the relevant spectral curve $\Sigma$. 
This way the calculation reduces to a zero modes counting on $\Sigma$. 
The exact result is then obtained by taking into account the redundancy
of parameters in the plane curve representation of the 
spectral curve $\Sigma$.

Linearizing the Hitchin equations~(\ref{insteq1}),~(\ref{insteq2}) gives 
\be
&&D_w\delta A_{\bar w}-D_{\bar w}\delta A_w-
ig^2\left[\bar X,\delta X\right]
+ig^2\left[X,\delta\bar X\right]=0\label{linear1}\\
&& D_w\delta X+i\left[\delta A_w,X\right]=0\, ,
\quad
D_{\bar w}\delta\bar X+i\left[\delta A_{\bar w},\bar X\right]=0\, .
\label{linear2}
\ee

The condition that tangent vectors be orthogonal to gauge transformations 
can be written in the form
\be
D_w\delta A_{\bar w}+D_{\bar w}\delta A_w+ig^2\left[\bar X,\delta X\right]
+ig^2\left[X,\delta\bar X\right]=0
\label{orth}
\ee
by making use of the scalar product
\be
<(\delta_1A,\delta_1X)|(\delta_2A,\delta_2X)>=
\int_{Cyl}d^2w {\rm Tr}\left[
\delta_1A_w\delta_2A_{\bar w} + g^2\delta_1X \delta_2\bar X +\quad {\rm h.c.}
\right].\label{scprod}
\ee

Let us now take the $g\to\infty$ limit of the above equations. 
The limit background, labeled with $\infty$, is then a 
strong coupling limit instanton characterized by a definite spectral curve 
$\Sigma$, as explained in \cite{bbn2} and in section~\ref{2} above,
\be
A^\infty_w=-iU\partial_wU^+\, , \quad X^\infty=U\hat X U^+
\label{back}\ee
and the equations for the tangent space to these instantons become
the following ones
\be
&&D^\infty_{\bar w}\delta A_w=0\, ,
\label{uno}\\
&&\left[\bar X^\infty,\delta X\right]=0,\label{due}\\
&&D^\infty_w\delta X+i\left[\delta A_w,X^\infty\right]=0\, ,\label{tre}
\ee
together their hermitian conjugates. We specify that these equations
characterize the tangent space to the stringy instantons (see comment
at the end of section~\ref{3.2}), not necessarily to the most general
Hitchin solutions.  To solve these equations we use the lifting
tecnique we exploited in \cite{bbn2}, which we refer to for notation
and some technical points.

Let ${\tt t}$ be the Cartan subalgebra in $u(N)$ obtained as $U{\tt t}_dU^+$
where ${\tt t}_d$ is the diagonal one.
Since $\bar X^\infty\in {\tt t}$, then, by~(\ref{due}), 
also $\delta X\in {\tt t}$.
Using~(\ref{back}) it is easy to see that also
$D_w^\infty\delta X\in {\tt t}$ and, by~(\ref{tre}) and $X\in {\tt t}$,
we conclude that $\delta A_w\in {\tt t}$.

This means that all the variations are in the Cartan subalgebra 
${\tt t}$, and the above equations reduce to
\be
D^\infty_{\bar w}\delta A_w=0
\, ,\quad
D^\infty_w\delta X=0 \quad {\rm where }
\quad \delta A_w\, ,\,\, \delta X\in{\tt t}\,,
\label{quattro}\ee
plus their hermitian conjugates.

To count the solutions of eq.~(\ref{quattro}), we lift them to the
spectral surface defined by the limit background field
(\ref{back}). They can be shown to reduce to
\be
\partial_{\bar z} {\delta \tilde A}_z=0
\, ,\quad
\partial_z {\delta \tilde X}=0\,,
\label{cinque}\ee
where ${\delta \tilde A}_z d\, z$ is therefore a holomorphic
differential on $\Sigma$ and ${\delta \tilde X}$ a holomorphic scalar.
Following the doubling trick explained in \cite{bbn2} we get a total
number of solutions equal to $\hat h +1=(2h+n-1)+1=2h+n$.

Finally, from this number we have to subtract 3; in fact the 3 parameters
of the transformations
of the variable $y$ belonging to ${\cal H}$ (section~\ref{4.3}) appear as genuine
moduli in the above counting, while, of course, they are not.

\appendix

\section{Singularities of plane curves}

Singularities of plane curves are a classical subject. Here we briefly review
the most useful methods for their study, which we partially use in the text.

Let us denote the curve defined by the polynomial $P(y, z)=0$ with
$\Sigma$. A {\it singular} point is a point of $\Sigma$, in which
$\partial_y P$ and $\partial_z P$ vanish.  Suppose, without loss of
generality, that the point we are interested in $\{y=0, z=0\}$.  Let
us find the local behaviour of the solutions $y(z)$ around it. This
can be obtained from the {\it Puiseux expansion}, which in turn can be
reconstructed from the Newton polygon.  A Newton polygon may in
general have several sides: each side represents (locally) a factor of
the curve, which can in turn be reducible, depending on the number of
points of the lattice that lie on it. Consider now the steepest of
these sides near the origin: call it $L_0$, and $\mu_0$ its
slope. $L_0$ defines a grading according to which $y$ has the same
weight as $z^{\mu_0}$. Now we approximate the solution of the equation
by the ansatz $y=t_0 z^{\mu_0}$. If we insert this in $P$, the latter
takes the form $P= z^{\mu_0} g(t_0)$ plus other terms which are higher
order terms with respect to the grading defined by $L_0$; $g $ is a
polynomial in $t_0$. Solving $g(t_0)=0$ we find in general several
values of $t_0$.  For each of them we can now substitute $y= t_0
(z^{\mu_0}+ y_1)$ in $P$ and find a new polynomial $P_1$, together
with a new Newton polygon.  Looking at its steepest side, say $L_1$,
with slope $\mu_1$, we can try a new solution in the form $y_1= t_1
z^{\mu_1}$, and so on.

This procedure needs not come to an end after a finite number of steps;
however, the series 
\be
y= t_0 (z^{\mu_0}+ t_1 (z^{\mu_1}+ \dots))\label{puis}
\ee
can be shown to converge to the solutions of $F$ in a sufficiently
small neighborhood of the origin. The numbers $\mu_0, \mu_1,...$ are
rational and increasing. A series like~(\ref{puis}) is called a {\it
Puiseux expansion} for $P$.

Puiseux expansions can now be used to analyse how solutions 'twist
around one anoth\-er', and to classify singularities up to topological
equivalence; around $z=0$ and $z=\infty$ this leads to nothing else
than the string interpretation already discussed in the text -
although, from a mathematical point of view, there is something more
to it.  Consider the first term $\mu_0$ of the Puiseux expansion and
suppose $\mu_0>0$.  Let us write $y^{q_0}= z^{p_0}$ ($\mu_0
=p_0/q_0$), and let us follow the solutions $y$ as $z= \epsilon
e^{2\pi i \phi}$ makes a small loop around zero. They lie on a torus
$\{|y|= \epsilon^{\mu_0}$, $|z| = \epsilon\}$, and wind around its two
cycles $q_0$ and $p_0$ times respectively. We can represent the union
of various circles covered by the solutions in $R^3$, and it turns out
to be a link of knots.  We can simply represent such knots by first
drawing a braid with $q_0$ threads which permute $p_0$ times, and then
connect initial with final points, as in figure~\ref{knot}.

\EPSFIGURE{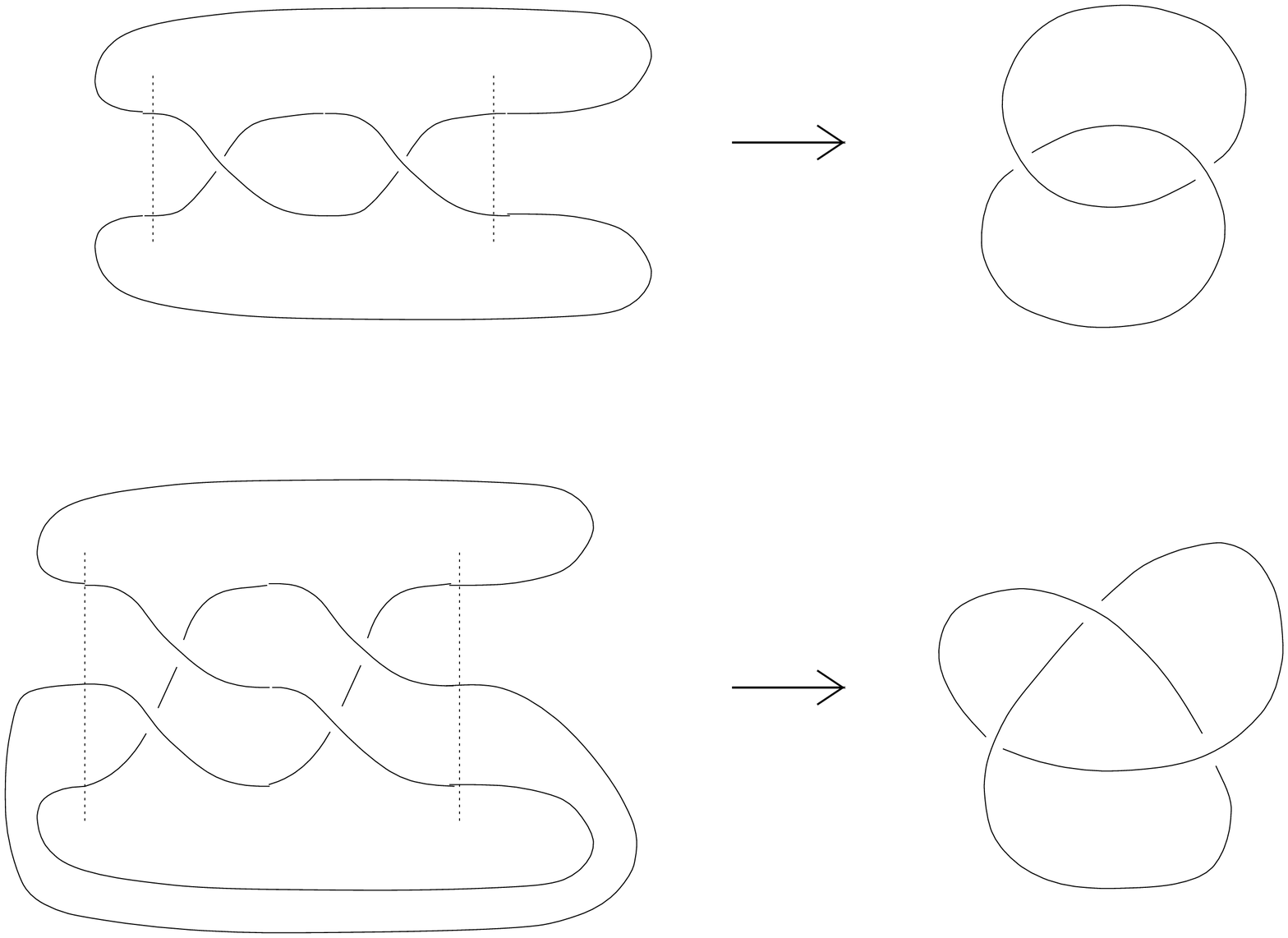,width=18em} {Knots corresponding to $y^2=z^2$ and
$y^3= z^2$ singularities.\label{knot}}

This knot is not in general the one that describes our singularity;
one has to take into account the higher terms in the expansion. If
however $\mu_1 =p_1/q_1$ and $q_1=q_0$, this does not really change
the braid (nor the knot).  If it is not so, one is led to a more
complicated figure, a knot which winds around a little torus
constructed around the previously defined knot; this is called an {\it
iterated torus knot}. This process can be shown to stop - in fact in
our case we know that the number of threads of the braid really is $N$
- and the topological type of the knot can be completely encoded in a
finite ordered set of pairs, of which the first is $\{p_0,q_0 \}$,
called {\it Puiseux pairs}. The knot in turn classifies the
singularity; it is in fact homeomorphic to the intersection of the
curve with a suitably small sphere $S^3$.

This latter fact means that the knot which classifies the singularity
is in fact the projection of the physical string embedded in the
target space, projected down to $3+1$ dimensions. What is really of
interest to us is the number of component, since all knots and links
can be undone in higher dimensions; and this is readily found also
without drawing knots.

In relation with our discussion in the text, perhaps the most useful
thing one can explore, is the contribution of a given singularity to
the genus. For singular curves the genus is defined as that of their
{\it resolution}; given $\Sigma$ and the set ${\cal S}$ of its
singular points, a resolution of $\Sigma$ is a smooth curve $\tilde
\Sigma$, (usually embedded in a larger space than the original curve),
together with a holomorphic projection $p: \ \tilde \Sigma
\rightarrow \Sigma$, such that its restriction $\tilde p: \
\tilde\Sigma - p^{-1}({\cal S}) \rightarrow \Sigma - {\cal S}$ is a
biholomorphism.

In words, a resolution can be locally achieved by replacing the
singular point by some space; Puiseux expansions are resolutions if we
replace a singular point with all possible behaviours near it. But
this is not a very economical resolution since the resulting space is
huge.  A handier way is to replace the singular point by a sphere -
called {\it exceptional divisor}. This is precisely the well known
procedure of blowing up the singularity, which in addition has the
advantage of exhibiting the smooth curve in some ${\CP^k}$.

Fortunately one needs not compute explicitely the blown-up curve;
every singular point simply gives a contribution that can be computed
as follows.  If we blow up the singularity, we get an exceptional
divisor, which meets the {\it strict preimage} $\tilde \Sigma -
p^{-1}({\cal S})$ of the curve in a number of points - called {\it
infinitely near}; some of these can be singular again, and we can blow
them up in turn, getting new infinitely near points; it can be shown
that this procedure eventually comes to an end.  Now, if $h$ is the
genus of $\tilde \Sigma$, the genus of the original curve $\Sigma$ is
given by
\begin{equation}
h- {1\over 2}\sum_k \nu_k ( \nu_k -1 )\,,
\end{equation}
where $k$ runs over all the infinitely near points $P_k$ of the
singular point and $\nu_k$ are their multiplicity, i.e.\ is the order
of the first non-vanishing term of the Taylor expansion of $P(y,z)$ at
$P_k$.

\acknowledgments 

We would like to acknowledge valuable discussions we had with E.
Aldrovandi, M. Asorey, M. Bertola, M. Bianchi, M. Bochicchio,
M. Caselle, A. D'Adda, B.Dubrovin, G. Falqui, T. Grava, C. Reina and
A. Zampa. One of us (L. B.) would like to thank ESI, Wien, for the
kind hospitality extended to him during the elaboration of this work.
This research was partially supported by EC TMR Programme,
grant FMRX-CT96-0012, and by the Italian MURST for the program
``Fisica Teorica delle Interazioni Fondamentali''.

\end{document}